\documentstyle[twocolumn,aps,prb,epsf]{revtex}
\begin{document} 
\title{Pairing correlation in the two- and three-leg Hubbard ladders \\
--- Renormalization and quantum Monte Carlo studies} 
\author{Takashi Kimura*,
Kazuhiko Kuroki, and Hideo Aoki} 
\address{Department of Physics, University of Tokyo, 
Hongo, Tokyo 113, Japan\\} 
\date{\today}
\maketitle

\begin{abstract} 
In order to shed light whether the `even-odd conjecture' 
(even numbers of legs will superconduct accompanied by a spin 
gap while odd ones do not) for correlated electrons in ladder 
systems, 
the pairing correlation is studied for the Hubbard model 
on a two- and three-leg ladders.   
We have employed both the weak-coupling renormalization group 
and the quantum Monte Carlo (QMC) method for strong interactions. 
For the two-leg Hubbard ladder, 
a systematic QMC (with a controlled level spacings) has 
detected an enhanced pairing correlation, 
which is consistent with the weak-coupling prediction. 
We also calculate the correlation functions in
the three-leg Hubbard ladder and show that 
the weak-coupling study predicts the dominant superconductivity, 
which refutes the naive even-odd conjecture.  
A crucial point is a spin gap for only some of the multiple 
spin modes is enough to make the ladder superconduct 
with a pairing symmetry (d-like here) compatible with the gapped mode.  
A QMC study for the three-leg ladder 
endorses the enhanced pairing correlation. 
\end{abstract}
\section{Introduction}
Over the past several years, strongly correlated electron 
systems with quasi-one-dimensional (1D) ladder structures 
have received much attention theoretically and experimentally. 
Experimental studies have received much impetus, since 
cuprate compounds containing such structures have been fabricated 
recently.\cite{DagRice} 

The idea was inspired theoretically in 1986, 
when Schulz\cite{SchulzAF} conjectured the following.  
If we consider a gas of repulsively interacting electrons 
on a ladder, the undoped system will be 
a Mott insulator, so that we may consider the system as 
an $S=1/2$ antiferromagnetic (AF) Heisenberg magnet on a ladder. 
Then an AF ladder with $N$-legs should be 
similar to a AF $S=N/2$ single chain, which is exactly 
Haldane's system.\cite{SchulzAF,Haldane,Nishiyama} 
For the latter Haldane's conjecture predicts that 
the spin excitation should be gapless for a half-odd-integer spin 
($N$: odd) 
or gapful for an integer spin ($N$: even).  If the situation would be similar 
in ladders, a ladder having an even number of legs 
will have a spin gap, which should indicate 
that the ground state is a `spin liquid' where the quantum 
fluctuation is so large that spins cannot order.  On the other hand 
an odd number of legs 
will have gapless spin excitations, which should indicate that 
the ground state has an AF order.  
The presence of a spin gap in the former case may be a good news for 
superconductivity, since, there is a body of ideas 
dictating that a way to obtain superconductivity is to 
carrier-dope a system that has a spin gap in the course of the 
study of high-$T_C$ superconductivity.  
Such a scenario has been put forward by Rice {\it et al.}\cite{Rice}

As far as the spin gap is concerned, 
both theoretical\cite{Dagotto,White,Greven,Poilblanc3,Hatano} 
and experimental\cite{Azuma,Ishida1,Kojima,Hiroi2} studies 
on the undoped-ladder systems have indeed supported the conjecture.

The spin-gap conjecture has recently been confirmed 
experimentally\cite{Azuma,Ishida1,Kojima}.  
Namely, a class of cuprates, Sr$_{n-1}$Cu$_{n}$O$_{2n-1}$, 
has $n$-leg ladders on a CuO$_2$ plane, and 
the two-leg ladders in SrCu$_2$O$_3$ exhibit a spin-liquid 
behavior characteristic of finite spin-correlation lengths, 
while the three-leg ladders in Sr$_2$Cu$_3$O$_5$ have an AF behavior. 

Rice {\it et al.} have further conjectured for doped systems that  
an even-numbered ladder should have a dominant 
interchain {\it d-wave-like} pairing correlation as expected from 
the persistent spin gap away from half-filling.\cite{Rice}  
The conjecture is partly based on an exact diagonalization study
for finite systems for a two-leg $t-J$ ladder 
by Dagotto {\it et al.}\cite{Dagotto} 
This was then followed by analytical\cite{Sigrist}
and numerical\cite{Poilblanc3,Tsunetsugu,Hayward1,Hayward2,Hayward3,Sano}
works on the doped $t-J$ ladder, 
which support
the dominant pairing correlation in a certain region. 
In the phase diagram, the region for 
the dominant pairing
correlation appears at lower values of exchange coupling $J$ 
than in the case of a single chain. 
Experimentally Uehara {\it et al.}\cite{Uehara} have recently observed 
superconductivity in a two-leg ladder material 
Sr$_{0.4}$Ca$_{13.6}$Cu$_{24}$O$_{41.84}$ 
under high pressures. 

The Hubbard models on ladders are also of interest, 
since the Hubbard model may be regarded as an effective model 
for cuprates (Fig.1). 
Since there is no exact solution for the Hubbard ladder, 
a most reliable analytical method 
at present is the weak-coupling theory,\cite{Review,weak}
which, in the continuum limit, 
linearizes the band structure around the Fermi points 
to treat the interaction with a perturbative renormalization group. 

\begin{figure}
  \begin{center}
\leavevmode\epsfysize=20mm \epsfbox{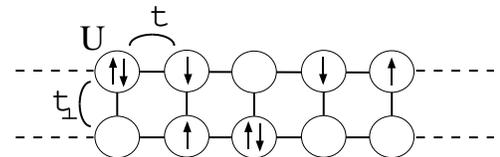}
  \caption{Two-leg Hubbard ladder model; 
$t(t_\perp)$ and $U$ are the intrachain(interchain) hopping 
and the on-site interaction, respectively.}
  \label{fig:hubbard2}
  \end{center}
\end{figure}

The weak-coupling theory has been 
applied to the two-leg Hubbard ladder.
\cite{Balents,Fabrizio,Nagaosa,Schulz2} 
At half-filling, the system reduces to a spin-liquid 
insulator having both charge and spin gaps\cite{Balents} 
with a finite SDW correlation length. 
Although one might expect that the Hubbard ladder would not exhibit a 
sizeable spin gap at half-filling
unlike the $t-J$ ladder, 
the spin gap for the Hubbard model estimated with DMRG 
by Noack {\it et al.}\cite{Noack1,Noack3} is as large as 0.13$t
(\sim 400$K for $t\sim 0.3$eV for $U=8t$ with $t_{\perp}=t$, 
which should correspond to the cuprates).  
The magnitude
of the spin gap is comparable with the spin gap ($\sim$ 400K) experimentally
estimated from the magnitude of susceptibility for SrCu$_2$O$_3$.

When the carrier is doped, the weak-coupling theory supports 
the dominance of the pairing correlation whose symmetry is 
the same as that of the $t-J$ ladder. 
The relevant scattering processes 
at the fixed point in the renormalization flow 
are the pair-tunneling process across 
the bonding and the anti-bonding bands (Fig.2), and the 
backward-scattering process within each band. 
The importance of the pair-tunneling 
across the two bands (which exists in two or larger numbers of legs)
for the dominance of pairing correlation in the two-leg Hubbard ladder
is reminiscent of the Suhl-Kondo mechanism, that 
was proposed back in the 1950's for superconductivity
in the transition metals with two (s- and d-like) bands.\cite{Suhl,Kondo} 
Muttalib and Emery
have shown another example of the pair-tunneling mechanism for 
superconductivity with purely repulsive interactions.\cite{Muttalib} 
More recently, the superconductivity in $t-t^{\prime}-U$ model,
which may be relevant  for the chains that alternate with the 
ladder layers in the cuprates\cite{Uchida},
has also been studied analytically\cite{Fabrizio2} 
and numerically\cite{Arita} as a 1D ladder-like system. 

\begin{figure}
  \begin{center}
    \leavevmode\epsfysize=50mm \epsfbox{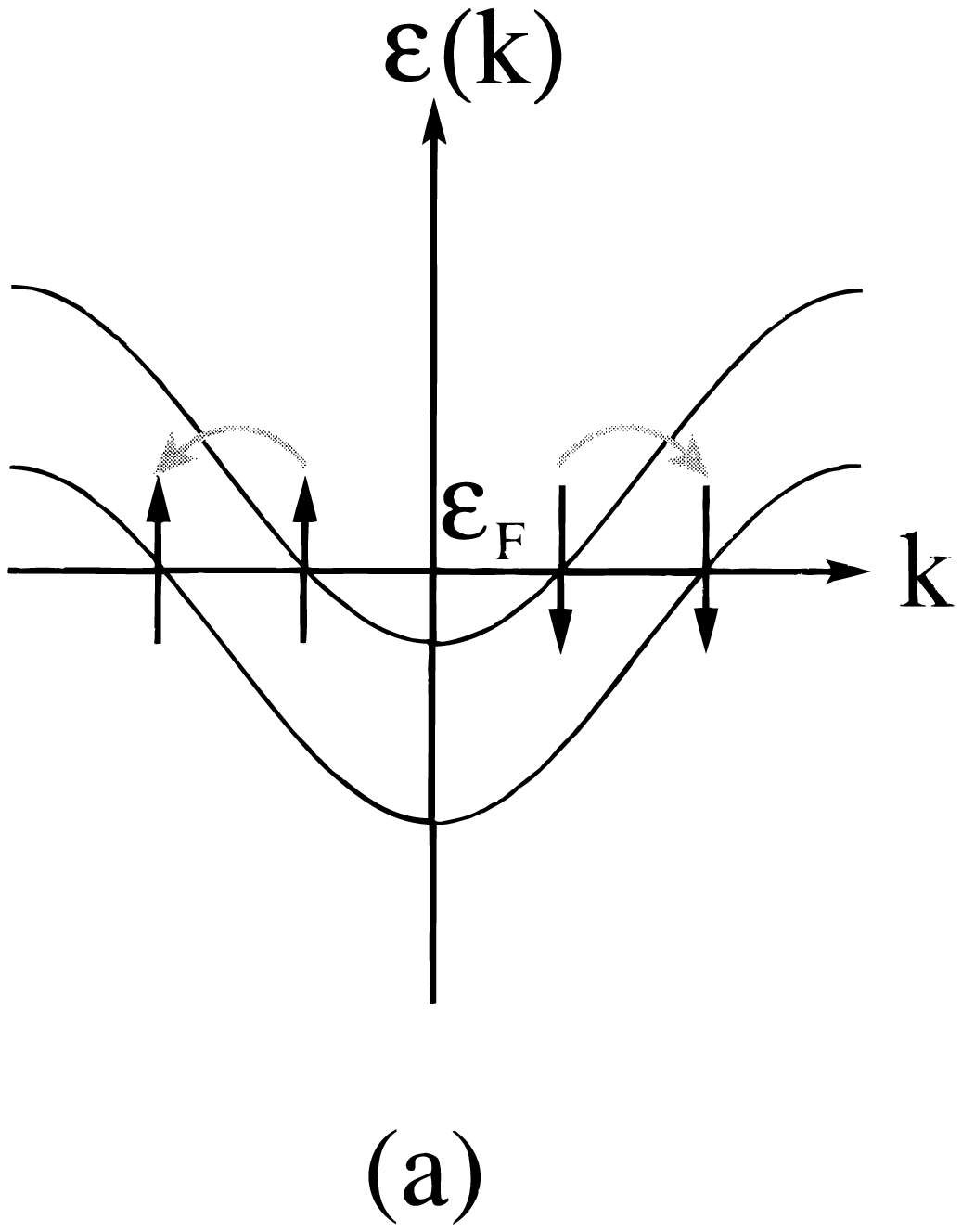}
     \leavevmode\epsfysize=50mm \epsfbox{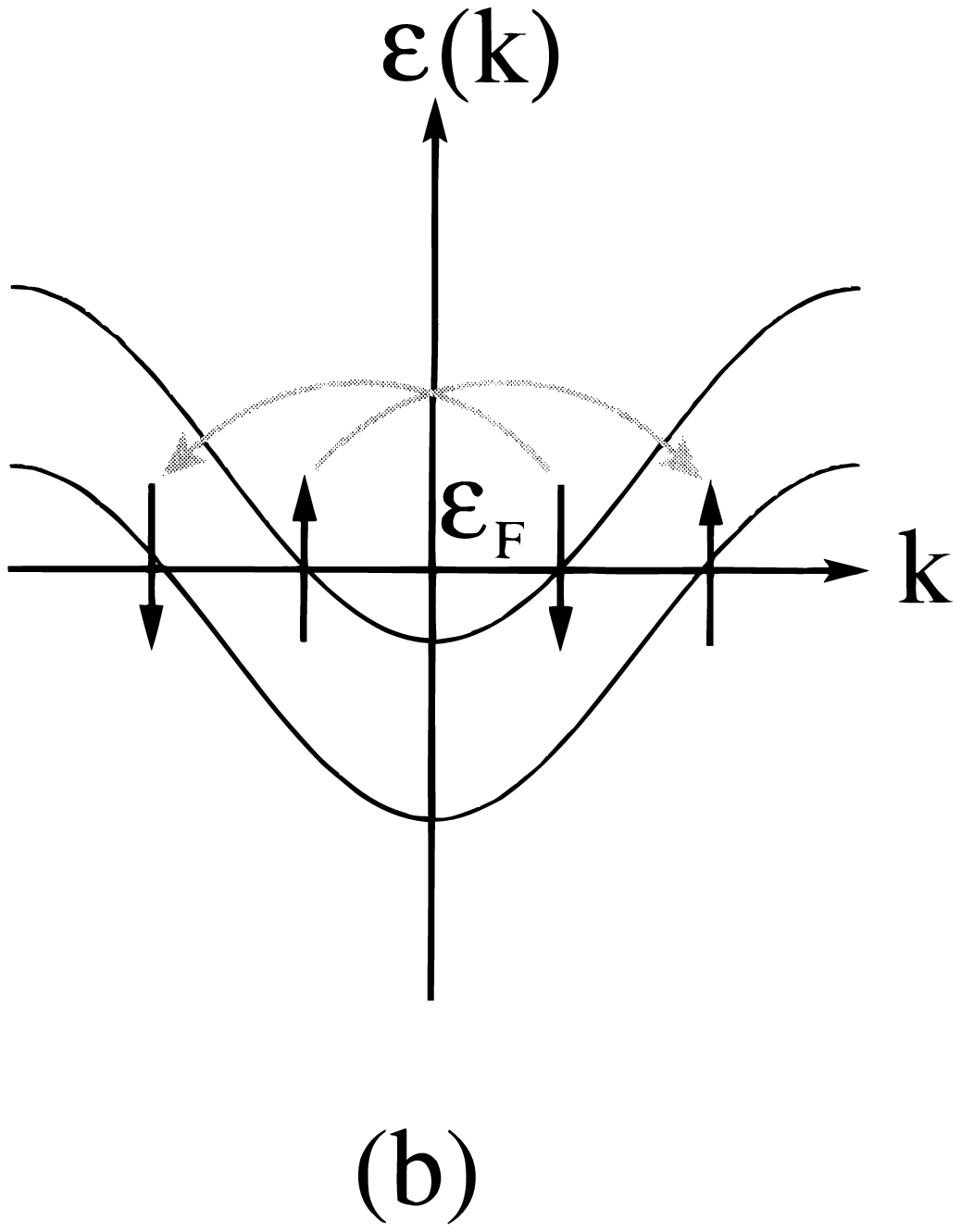}
  \caption{Relevant pair-tunneling processes in two-leg Hubbard ladder;
Fig.(a) (Fig.(b)) is the forward (backward) type pair-tunneling process.}
  \end{center}
  \label{fig:twolegb}
\end{figure}

The properties of the weak-coupling Hubbard ladder are similar to 
those of the $t-J$ ladder for the regime where the pairing correlation 
is dominant: in addition to the existence of the spin gap, 
the {\it duality relation}\cite{Nagaosa,Schulz2}, which suggests that 
the exponent for the pairing correlation 
$(\sim r^{1/(2K_{\rho}}))$ 
should be reciprocal to that for the $4k_F$ CDW 
correlation $(\sim r^{(2K_{\rho}}))$, holds 
in both the weak-coupling Hubbard ladder and the 
$t-J$ ladder.\cite{Hayward1,Hayward2,Sano} 
Here $K_\rho$ is the critical exponent for the gapless charge mode, 
which tends to unity in the weak-coupling limit. 
The similarity may come from the form of the excitation gaps 
in the bosonization description.  In both 
$t-J$ and 
Hubbard ladders\cite{Nagaosa},
the only gapless mode is a charge mode 
with none of the spin modes being gapless 
(`C1S0' phase in the language of the weak-coupling theory\cite{Balents}). 

However, there is a serious reservation for the weak-coupling theory. 
First, whether the model in the continuous limit is indeed 
equivalent to the lattice model is not obvious. 
More serious is the problem that the perturbational renormalization group 
is guaranteed to be valid only for an infinitesimally 
small interaction strengths in principle. 
Specifically, when there is a gap in the excitation, 
the renormalization flows into a strong-coupling regime, 
so that the perturbation theory may well break down even for small 
interaction strengths. 
A way to check the reliability of the weak-coupling theory 
is to treat finite systems with larger $U$ with numerical calculations 
such as  exact diagonalization, density-matrix renormalization group (DMRG), 
or  quantum Monte Carlo (QMC) methods\cite{Hirsch,MonteCarlo}. 
Numerical calculations, on the other hand, have drawbacks 
due to finite-size effects. 
Thus the weak-coupling theory and the numerical 
methods should be considered as being complementary. 

Specifically, the dominance of the pairing correlation 
is indeed a subtle problem in numerical calculations.  
Existing numerical results\cite{Noack1,Noack3,Yamaji,Asai,Noack2} 
do appear to be controversial, where 
some of the results are inconsistent with the weak-coupling prediction 
as detailed in Sec.II.

If we ignore these controversies, 
most of the existing theories support the dominance of 
the pairing correlation in the doped two-leg ladders. 
Then, an even more important unresolved problem for superconductivity 
in the doped ladders is the `even-odd' conjecture. 
One can naively expect that 
the absence of spin gaps in odd-numbered legs will signify an absence of 
dominating pairing correlation, 
which will end up with an `even-odd conjecture for superconductivity'. 
Indeed there has not been 
works looking into the pairing 
correlation functions for the three-leg ladder, which is the 
simplest realization of odd-numbered legs. 
White {\it et al.}\cite{WS} have studied two holes doped in 
the $t-J$ ladders at half-filling to find that two holes 
are bound in even-numbered (two or four) legs, while they are not 
in odd-numbered (three or five) legs, but (i)the existence of a 
binding energy for two carriers is not directly connected with 
the occurrence of  superconductivity, and (ii)the $t$-$J$ model 
and the Hubbard model may exhibit different behaviors. 

Thus the second purpose of the present paper 
is to study the pairing correlation 
in Hubbard ladder models with three legs as compared with the 
two-leg case. 

The organization of the present paper is as follows. 
We first study the two-leg Hubbard ladder model\cite{Kuroki} 
having intermediate 
interaction strengths with the QMC method in Sec.II. 
A new ingredient in this study is that 
we pay a special attention to the non-interacting ($U=0$) 
single-particle energy levels which are discrete 
in finite systems.  
We have found that if we make the levels on the two bands aligned 
(within an energy that is smaller than the spin gap) 
to mimic the thermodynamic limit, 
the weak-coupling result 
(an enhanced pairing correlation in the present case) is 
in fact reproduced for intermediate $U$. 
Further we check the effects of the inter- and intra-band  
Umklapp-scattering processes, which are 
expected to be present 
at the special band fillings
from the weak-coupling theory\cite{Balents}. 

In Sec.III we study three(odd)-leg Hubbard ladder model 
(Fig.3) 
with the weak-coupling theory\cite{Takashi1} 
using the enumeration of 
gapless modes by Arrigoni.\cite{Arrigoni} 
The system has one gapless spin mode and two gapful spin modes. 
Thus the gapful and the gapless spin modes coexist. 
The existence of the gapful modes is a result of the relevant interband 
pair-tunneling process across the top and the bottom bands (Fig.4). 
We find that this gives rise to 
a dominant pairing correlation 
across the central and the edge chains reflecting the gapful spin modes, 
which coexists with the subdominant but power-law decaying 
SDW correlation reflecting the gapless spin mode. 
Thus the Suhl-Kondo-like mechanism for superconductivity 
exists not only in two-leg systems 
but also in a three(odd)-leg system as well, 
while the SDW correlation also survives as expected. 
Schulz\cite{Schulz3} also found similar results independently.

\begin{figure}
  \begin{center}
\leavevmode\epsfysize=30mm \epsfbox{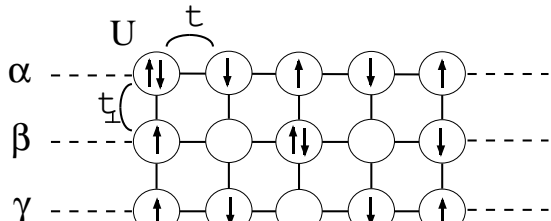}
  \caption{Three-leg Hubbard ladder model; 
$t(t_\perp)$ and $U$ are the intrachain (interchain) hopping 
and the on-site interaction, respectively.}
  \end{center}
  \label{fig:hubbard3}
\end{figure}

\begin{figure}
  \begin{center}
\leavevmode\epsfysize=50mm \epsfbox{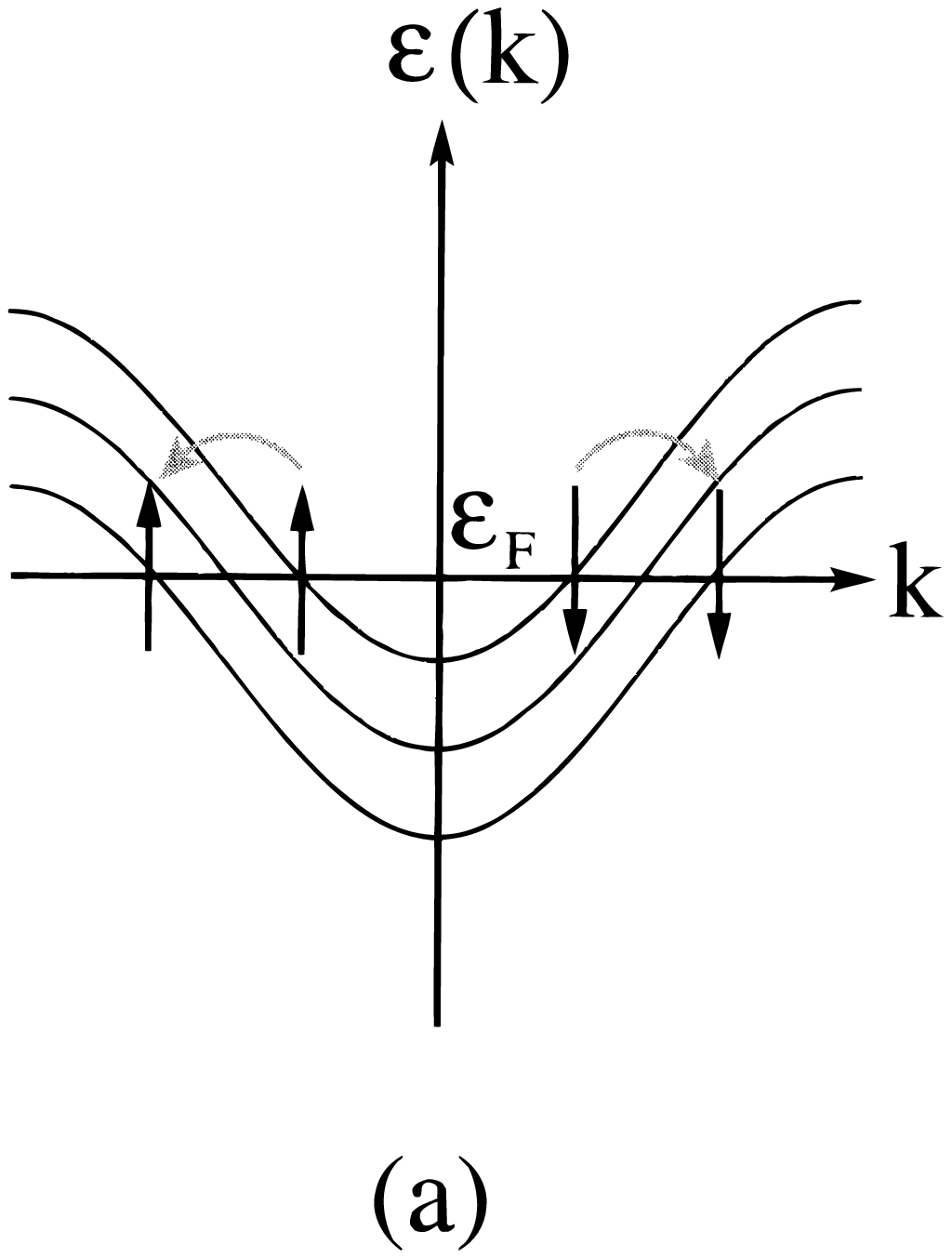}
\leavevmode\epsfysize=50mm \epsfbox{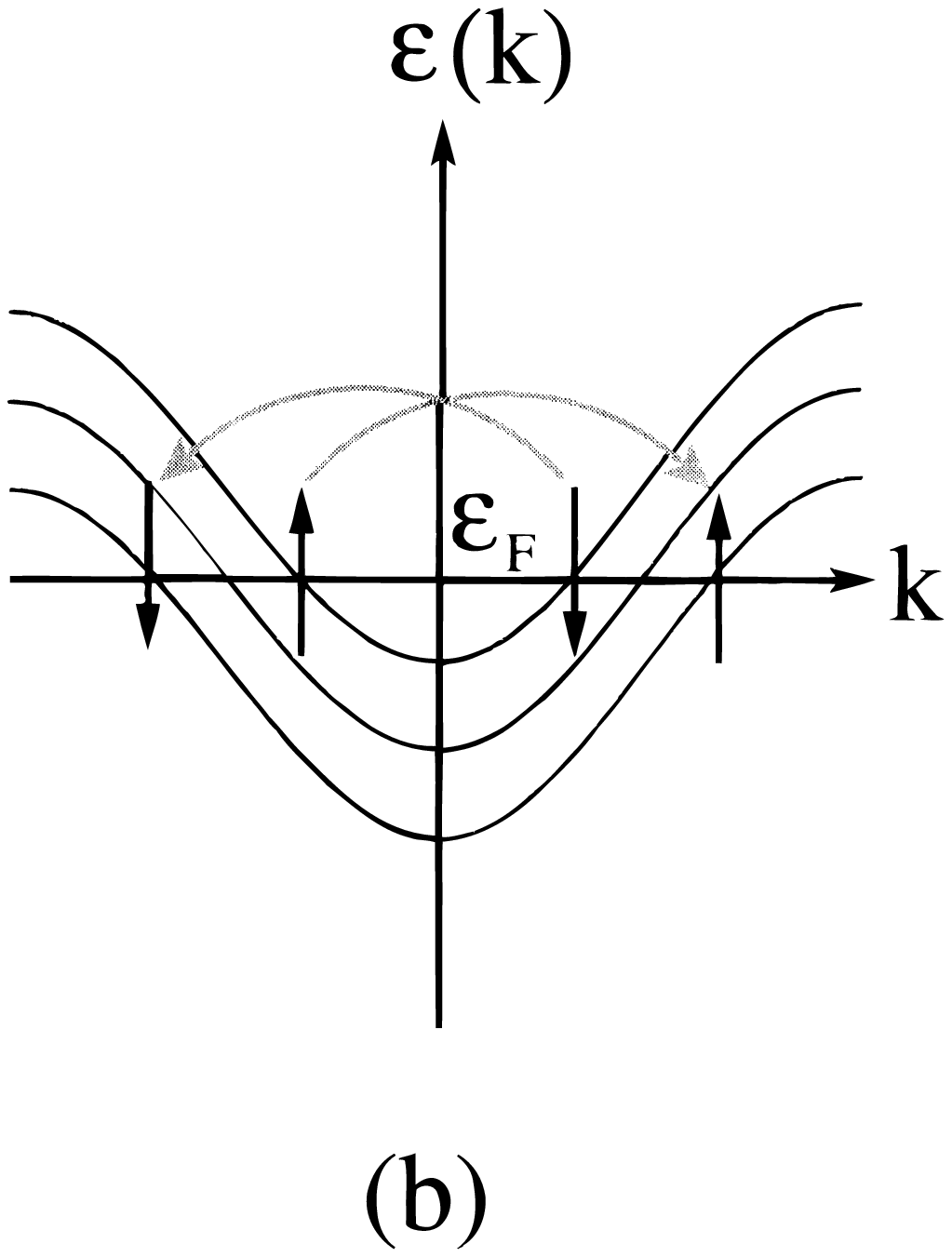}
     \caption{Relevant pair-tunneling processes 
in the three-leg Hubbard ladder;
Fig.(a) (Fig.(b)) is the forward (backward) type pair-tunneling process.}
  \end{center}
  \label{fig:threelegb}
\end{figure}

We then study the three-leg Hubbard ladder model in Sec.IV 
with a QMC calculation\cite{Takashi2} as in the two-leg case. 
The technique to detect the enhanced pairing correlation in the 
two-leg case is also valid in the three-leg case.
We found that the enhancement of the pairing correlation persists 
for the intermediate interaction strengths. 
We also study the effects of the Umklapp processes at special 
band fillings as in the two-leg case. 

The above numerical calculations for two or three legs 
are performed in a condition that the one-body energy levels are 
close to each other. We believe this condition mimics the situation
in the thermodynamic limit. In Sec.V, a circumstantial evidence for this
is given by a QMC calculation of the pairing correlation in 
the 1D attractive Hubbard model,
which can be exactly solved with the Bethe 
Ansatz\cite{Bethe}, so that the exact asymptotic form of the 
correlation functions and 
the values of the spin gap are known at arbitrary values of parameters. 
 
\section{Quantum Monte Carlo study of the pairing correlation 
in the two-leg Hubbard ladder} 

In this section, we study the pairing correlation 
for the two-leg Hubbard ladder (Fig.1). 
The Hamiltonian of the two-leg Hubbard ladder 
is given in standard notations as 
\begin{eqnarray} 
{\cal H}&=&-t\sum_{\alpha i \sigma} 
(c_{\alpha i\sigma}^\dagger c_{\alpha i+1\sigma}+{\rm h.c.})\nonumber\\&&
-t_{\perp}\sum_{i \sigma}
(c_{1,i\sigma}^\dagger c_{2,i\sigma}+{\rm h.c.})
+U\sum_{\alpha i} n_{\alpha i\uparrow}n_{\alpha i\downarrow},
\end{eqnarray}
where $\alpha(=1,2)$ specifies the chains and $t(t_{\perp})$ is 
the intra(inter)-chain hopping.

From an analytical point of view,
if the system is free from Umklapp processes, 
the weak-coupling theory with the bosonization combined with 
the renormalization-group 
techniques \cite{Balents,Fabrizio,Nagaosa,Schulz2} 
has indeed shown that the the two-leg Hubbard ladder has a spin gap 
and that the correlation function of the interchain 
{\it d-wave-like} pairing order parameter, 
$O_i=(c_{1i\uparrow} c_{2i\downarrow}-c_{1i\downarrow} c_{2i\uparrow})
/\sqrt{2}$, 
decays as $\sim r^{-1/(2K_{\rho})}$ 
as a result of the relevant pair-tunneling process (Fig.2),
where $K_{\rho}$ is the critical exponent for the total-charge-density mode 
being only gapless and tends to unity in the weak-coupling limit.

Since SDW and $2k_F$ CDW correlations have to decay 
exponentially in the presence of a spin gap in a two-leg ladder, 
the only phase competing with pairing correlation 
will be $4k_F$ CDW correlation, which should 
decay as $r^{-2K_{\rho}}$. Hence the pairing correlation dominates 
over all the others if $K_{\rho}>1/2$.                                                                   
In numerical calculations, however, the dominance of the pairing correlation 
in the Hubbard ladder appears to be a subtle problem. 
Namely, a DMRG study by Noack {\it et al.} for the doped Hubbard ladder  
with  $n=0.875$, $U/t=8$, and $t_{\perp}=t$ 
shows no enhancement of the pairing correlation 
over the $U=0$ result\cite{Noack1,Noack3}, while 
they do find an enhancement at $t_{\perp}=1.5t$\cite{Noack3,Noack2}. 
Asai performed a quantum Monte Carlo (QMC) calculation for a 
36 rungs ladder with $n=0.833$, $U/t=2$ and $t_{\perp}=1.5t$\cite{Asai}, 
in which no enhancement of the pairing correlation was found. 
On the other hand, Yamaji {\it et al.} have found an enhancement 
for the values of the parameters when the lowest anti-bonding band levels 
for $U=0$ approach the highest occupied bonding band levels, 
although their results have not been conclusive due to the small system sizes 
($\leq 6$ rungs).\cite{Yamaji}  
Thus, the existing analytical and numerical results appear to be 
controversial in the two-leg Hubbard ladder. 
                                                   
Another point is that the above results are obtained away from 
special fillings where the Umklapp-scattering processes are irrelevant. 
Recently, Balents and Fisher\cite{Balents} 
proposed a weak-coupling phase diagram 
(Fig.5) which displays 
the numbers of the gapless spin 
and charge phases on the $t_{\perp}-n$ plane 
(where $n$ is the band filling). 
The effects of the Umklapp processes are also discussed there. 
At half-filling the interband Umklapp processes become relevant  
resulting in a spin-liquid insulator in which the pairing 
correlation decays exponentially. 
In addition, the intraband Umklapp process within the bonding band 
becomes relevant resulting in a gap in one charge mode 
in a certain parameter region where the bonding band is reduced to 
a half-filled band. 
This phase is called `C1S2' phase because there are one gapless 
and two gapfull charge modes, while the phase at half-filling 
is called `C0S0' phase because there is no gapless phase. 
We can expect that the pairing correlation decays exponentially 
or is at least suppressed  reflecting the existence of the charge gap, 
although the direct calculation 
of the pairing correlation has not been done. 
Thus we also study the effects of the Umklapp processes in this section, 
keeping in mind the above weak-coupling results. 

\begin{figure}
  \begin{center}
    \leavevmode
\leavevmode\epsfysize=55mm \epsfbox{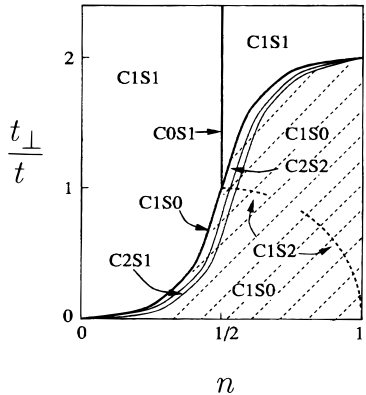}
  \caption{ Phase diagram in the weak-coupling limit ($U\rightarrow0$) 
given in ref. [39]; 
the numbers of the gapless 
charge and spin modes ($x$ and $y$, respectively) are denoted as C$x$S$y$ 
and $n$ is the band filling. 
In the dark region both of the two bands cross the Fermi level.
}
  \end{center}
  \label{fig:balents}
\end{figure}

In the remaining of this section, we perform an extensive projector QMC 
calculation\cite{Hirsch,MonteCarlo}
to investigate the ground state correlation function 
$P(r)\equiv \langle O_{i+r}^\dagger O_{i}\rangle$
for the pairing in the Hubbard ladder\cite{Kuroki}  
with $t_{\perp}\sim t$, especially in order to clarify the origin of 
the discrepancies among the existing results.  
We conclude that the discreteness of energy levels in finite systems 
affects the pairing correlation enormously. 

The details of the QMC calculation are the following.
We assumed the the periodic boundary conditions along 
the chain direction, $c_{N+1} \equiv c_{1}$ (where $N$ labels the rungs)
and took the non-interacting Fermi sea as the trial state 
The projection imaginary time $\tau$ was taken to be $\sim60/t$.
We need such a large $\tau$ to ensure the convergence of especially 
the long-range part of the pairing correlation.  
This sharply contrasts with the situation for single chains, 
where $\tau\sim 20/t$ suffices 
for the same sample length as considered here.  
The large value of $\tau$, along with a large 
on-site repulsion $U$, makes the negative-sign problem serious, 
so that the calculation is feasible for $U/t\leq 2$. 
In the Trotter decomposition, the imaginary time increment
$[\tau/$(number of Trotter slices)] is taken to be
$\leq 0.1$. 
We have concentrated on 
band fillings for which the closed-shell condition 
(no degeneracy in the non-interacting Fermi sea) is met.  
We set $t=1$ in the remaining of this section.

In the beginning we show in Fig.6 the result for $P(r)$
for $t_{\perp}=0.98$ and $t_{\perp}=1.03$ with $U=1$ and 
the band filling $n=0.867=52$ electrons/ (30 rungs $\times$ 2 legs).  
The $U=0$ result (dashed line) 
for these two values of $t_{\perp}$ are identical 
because the Fermi sea remains unchanged.
However, if we turn on $U$, the 5\% 
change in the $t_{\perp}=0.98\rightarrow1.03$
is enough to cause a dramatic
change in the pairing correlation: 
for $t_{\perp}=0.98$ the correlation has a large
enhancement over the $U=0$ result at large distances, while 
the enhancement is not seen for $t_{\perp}=1.03$.

\begin{figure}
   \begin{center}
\leavevmode\epsfysize=70mm \epsfbox{figure6.epsi}
   \caption{Pairing correlation function, $P(r)$, plotted 
against the real-space distance $r$ in a 30 rungs Hubbard 
ladder having 52 electrons for $U=1$ with $t_\perp =0.98$ ($\Box$)
and $t_{\perp}=1.03$ ($\diamond$ ). 
The dashed line is the noninteracting
result for the same system size, while the straight dotted line 
represents $\propto 1/r^2$. The solid line is a fit to the 
$U=1$ result with $t_\perp =0.98$ (see text).}
  \end{center}
  \label{fig:fig21a}
\end{figure}

\begin{figure}
   \begin{center}
\leavevmode\epsfysize=50mm \epsfbox{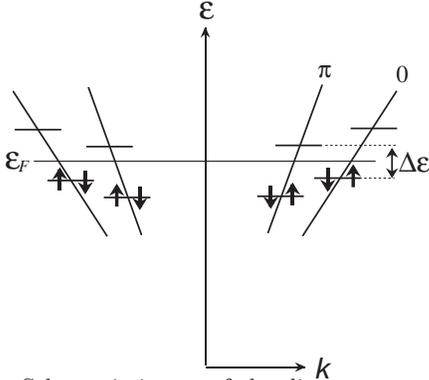}
   \caption{
Schematic image of the discrete energy levels of both 
bonding ($0$) and anti-bonding ($\pi$)
bands for $U=0$.}
  \end{center}
  \label{fig:fig21b}
\end{figure}

In fact we have deliberately chosen these values 
to control the alignment of the 
discrete energy levels at $U=0$. 
Namely, when $t_{\perp}=0.98$, 
the single-electron energy levels of the 
bonding and  anti-bonding bands for $U=0$ lie close to each other 
around the Fermi level with the 
level offset ($\Delta\varepsilon$ in Fig.7) 
being as small as 0.004, while
they are staggered for $t_{\perp}=1.03$ with the level offset of 0.1.
On the other hand, the size of the spin gap is known to be  
around $0.05$ for $U=8$,\cite{Noack2,Noack3} and is expected to be of the  
same order of magnitude or smaller for smaller values of $U$.
The present result then suggests that if the level offset 
$\Delta\varepsilon$ is too large compared to the spin gap 
(which should be O(0.01) for $U\sim t$,\cite{Noack3})  
the enhancement of the pairing correlation is smeared. 
By contrast, for a small enough $\Delta\varepsilon$, by which 
an infinite system is mimicked, the enhancement 
is indeed detected in agreement with the weak-coupling 
theory, in which the spin gap is assumed to 
be infinitely large at the fixed 
point of the renormalization flow. 
Similar situation is also found in the case of a QMC study of the 
pairing correlation in the $t-t^{\prime}-U$ model.\cite{Arita}

Our result is reminiscent of those obtained by 
Yamaji {\it et al.},\cite{Yamaji} 
who found an enhancement of the pairing correlation 
in a restricted parameter regime where the lowest anti-bonding 
levels approach the highest occupied bonding levels.
They conclude that the pairing correlation is dominant 
when the anti-bonding band `slightly touches' the Fermi level. 
However, our result in Fig.6 is obtained for the 
band filling for which no less than seven out of 30 anti-bonding levels 
are occupied at $U=0$.  
Hence the enhancement of the pairing correlation
is seen to be not restricted to the situation 
where the anti-bonding band edge touches the Fermi level. 

However, one might consider that the enhancement of the pairing correlation
in such a condition for the one-body energy levels should be rather 
due to finite size effects, although we believe the condition is 
generally relevant for bulk systems.
To clarify this point, we will further 
give a circumstantial evidence to justify that the 
condition for the one-body energy levels is relevant for bulk systems 
by a QMC study for 1D attractive Hubbard model in Sec.V.

Now, let us more closely look into the form of $P(r)$ for
$t_{\perp}=0.98$.  It is difficult to determine the exponent
from results for finite systems, but here we attempt 
to fit the 
data by assuming a trial function expected from the weak-coupling theory. 
Namely, we have fitted the data with the form,
\begin{eqnarray}
P(r)&=&\frac{1}{4\pi^2}\sum_{d=\pm}\{cr_d^{-1/2}\nonumber\\
&&+[(2-c)-\cos(2k_F^0 r_d)-\cos(2k_F^{\pi} r_d)]r_d^{-2}\}
\label{tri}
\end{eqnarray}
with the least-square fit (by taking logarithm of the data).
Because of the periodic boundary condition, 
we have to consider contributions from both ways around, so there are
two distances between the $0$-th and the $r$-th rung, i.e, $r_{+}=r$ and 
$r_{-}=N-r$.
The periods of the cosine terms are assumed to be the 
non-interacting Fermi wave numbers of the bonding and the anti-bonding 
bands in analogy with the single-chain case. 

The overall decay should be $1/r^2$ as in the pure 1D case 
in the weak-coupling limit.   
We have assumed the form $c/r^{1/2}$
as the dominant part of the correlation at large distances 
because this is what is expected in the weak-coupling theory.
Here $c$ is the only fitting parameter in the above trial function. 
A finite $U\sim 1$ may give some correction, but the result 
(solid line in Fig.6) fits to the numerical result 
surprisingly accurately with a best-fit $c=0.10$. 
If we least-square fit the exponent itself as 
$1/r^\alpha$, we have $\alpha<0.7$ with a similar accuracy.  
Thus a finite $U$ may change $\alpha$, but $\alpha>1$ may be excluded.  
To fit the short-range part of the data, 
a non-oscillating $(2-c)/r^2$ term is required, 
which is not present in the weak-coupling theory.  
We believe that this is because the weak-coupling theory only concerns with
the asymptotic form of the correlation functions.

In Fig.7, we show a result for a larger system size (42 rungs) 
for a slightly different electron density, $n=0.905$ with 76 electrons 
and $t_{\perp}=0.99$. We have again an excellent fit 
with $c=0.07$ this time.

\begin{figure}
   \begin{center}
\leavevmode\epsfysize=70mm \epsfbox{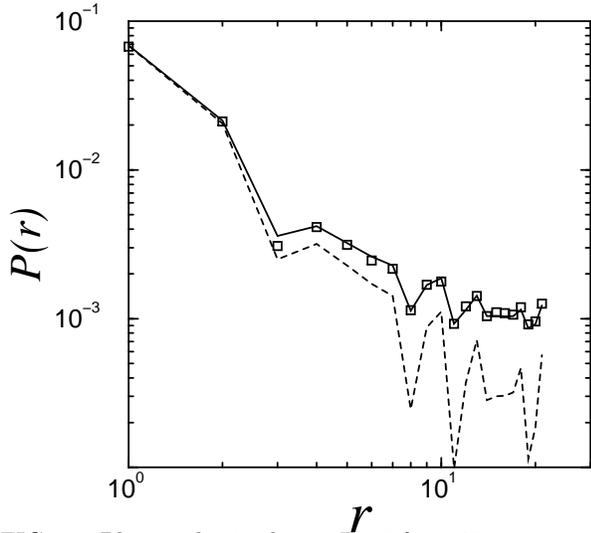}
      \caption{
Plot similar to that in Fig.6 for a 42 rungs system 
having 76 electrons with $t_{\perp}=0.99$.  
}
  \end{center}
  \label{fig:fig22}
\end{figure}

In Fig.8, we display the result for a larger $U=2$.  
We again have
a long-ranged $P(r)$ at large distances, 
although $P(r)$ is slightly reduced from 
the result for $U=1$.
This is consistent with the weak-coupling theory, in which
$K_\rho$ is a {\it decreasing} function of $U$ so that after
the spin gap opens for $U>0$, the pairing correlation decays 
faster for larger values of $U$.

\begin{figure}[htbp]
   \begin{center}
    \leavevmode
\leavevmode\epsfysize=70mm \epsfbox{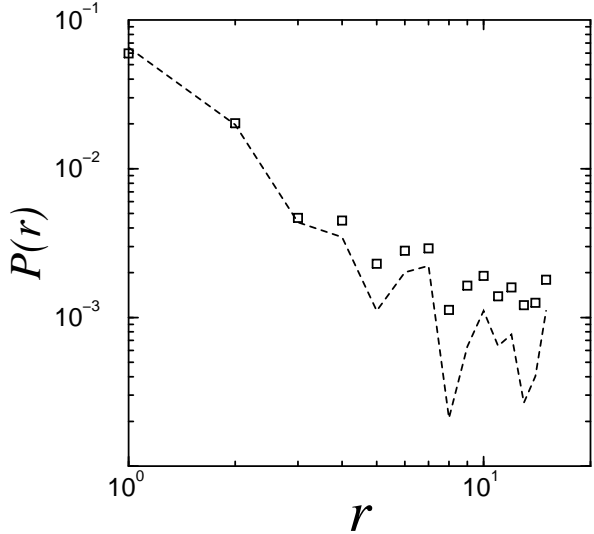}
     \caption{Plot similar to that in Fig.6
for $U=2$.
}
  \end{center}
  \label{fig:fig23}
\end{figure}

Now we explore the effects of the Umklapp processes.
For that purpose we concentrate on 
the filling dependence for a fixed interaction $U=2$.
We have tuned the value of $t_{\perp}$ to ensure that 
the level offset $(\Delta\varepsilon)$ at the Fermi level 
is as small as O($0.01$) for $U=0$.
In this way, we can single out the effects of the Umklapp processes from
those due to large values of $\Delta\varepsilon$.
If we first look at the half-filling (Fig.10), 
the decaying form is essentially similar to the $U=0$ result.  
At half-filling, the interband Umklapp processes emerge and, 
according to the weak-coupling 
theory, open a charge gap, which results in an exponential decay of the 
pairing correlation.
\noindent 
(We should note that there are two kinds of charge gaps. 
The one, which is produced by the pair-tunneling processes, 
causes the long range order of the Josephson phase resulting in 
the enhancement of the pairing correlation, 
while the other, which is produced by the Umklapp processes, 
causes the long range order of the phase of the CDW 
resulting in the suppression of the pairing correlation)

\begin{figure}
   \begin{center}
\leavevmode\epsfysize=70mm \epsfbox{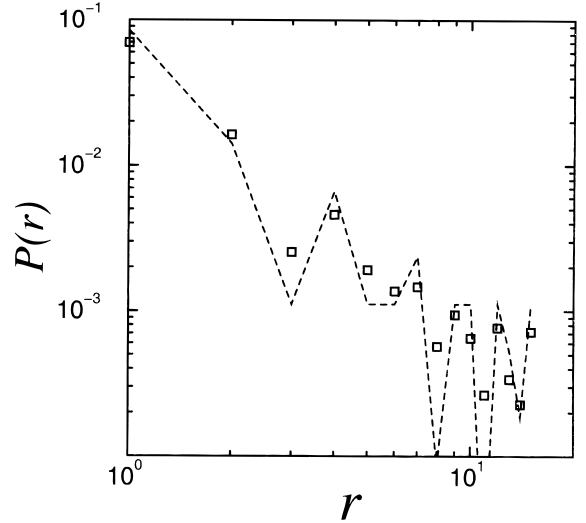}
      \caption{
Pairing correlation $P(r)$ ($\Box$) against $r$ 
for a 30 rungs system for $U=2$ with $t_{\perp}=0.99$
and 60 electrons (half-filled).
The dashed line represents the non-interacting result.
}
  \end{center}
  \label{fig:fig24a}

\end{figure}
\begin{figure}
   \begin{center}
\leavevmode\epsfysize=70mm \epsfbox{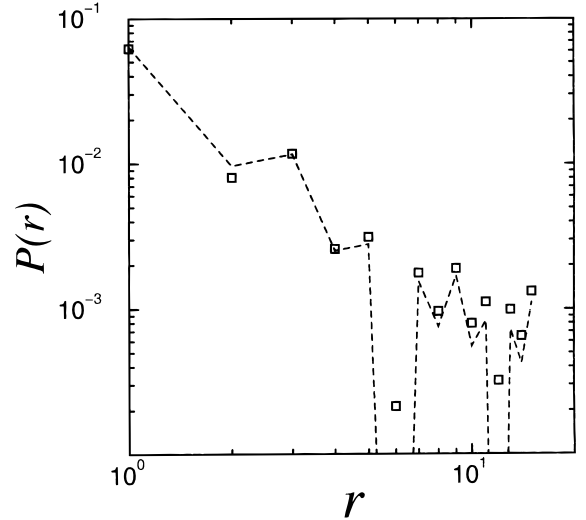}
     \caption{
Pairing correlation $P(r)$ ($\Box$) against $r$ 
for a 30 rungs system for $U=2$ with $t_{\perp}=1.01$ and 40 electrons 
(half-filled bonding band).
The dashed line represents the non-interacting result.
}  
  \end{center}
\label{fig:fig24b}
\end{figure}

It is difficult to tell from our data
whether $P(r)$ decays exponentially.  This is probably due to the 
smallness of the charge gap. In fact, 
the DMRG study by Noack {\it et al.}\cite{Noack1,Noack3} 
have detected an exponential decay for larger values of $U$,
for which a larger charge gap is expected.

When $n$ is decreased down to 0.667 (Fig.11), 
we again observe an absence of enhancement in $P(r)$. 
This is again consistent with the weak-coupling theory\cite{Balents}: 
for this band filling, the number of electrons in the bonding band 
coincides with $N(=30)$ at $U=0$, i.e., the bonding band is half-filled. 
This will then give rise to {\it intraband} Umklapp processes 
within the bonding band resulting in the `C1S2' phase 
as discussed in Section 2.1. The spin gap is 
destroyed and the singlet-pair of electrons or holes 
are prevented from forming, so that the pairing correlation 
will no longer decay slowly there.
Noack {\it et al.}\cite{Noack3} have suggested that the suppression of the 
spin gap and the pairing correlation function around $t_{\perp}=0.4t$ 
in ref.\cite{Noack2} may be due to the intraband Umklapp process.

In this section, we have detected the enhancement of the 
pairing correlation which is consistent with the weak-coupling theory. 

We have also seen that there are three possible 
causes that reduce the pairing correlation function in the 
Hubbard ladder:  

(i) the discreteness of the energy levels, 

(ii) reduction of $K_\rho$ for large values of $U/t$, and 

(iii) effect of intra- and interband Umklapp 
processes around specific band fillings.

The discreteness of the energy levels is a finite-size effect, 
while the others are present in infinite systems as well. 
We can make a possible interpretation for the existing results 
in terms of these effects.
For 60 electrons on 36 rungs with $t_{\perp}=1.5t$ 
in ref.\cite{Asai}, for instance, 
the non-interacting energy levels have a significant 
offset $\sim 0.15t$ between bonding and anti-bonding levels 
at the Fermi level, which may be the reason 
why the pairing correlation is not enhanced for $U/t=2$.  
For a large $U/t(=8)$ in ref.\cite{Noack1,Noack2,Noack3}, 
(ii) and/or (iii) in the above may possibly be important 
in making the pairing correlation for $t_{\perp}=t$ not enhanced.  
The effect (iii) should be more serious 
for $t_{\perp}=t$ than for $t_{\perp}=1.5t$ because the bonding 
band is closer to the half-filling in the former. On the other hand, 
the discreteness of the energy levels might exert some effects as well, 
since the non-interacting energy levels 
for a 32-rung ladder with 56 electrons $(n=0.875)$ 
in an open boundary condition have an offset 
of $0.15t$ at the Fermi level for 
$t_{\perp}=t$ while the offset is $0.03t$ for $t_{\perp}=1.5t$. 

Let us comment on a possible relevance of the present result 
to the superconductivity reported recently for a cuprate 
ladder\cite{Uehara}, especially for the 
pressure dependence. 
The material is Sr$_{0.4}$Ca$_{13.6}$Cu$_{24}$O$_{41.84}$, 
which contains layers consisting of 
two-leg ladders and those consisting of 1D chains.
Superconductivity is not observed in the ambient pressure, while 
it appears with 
$T_C\sim 10K$ under the pressure of 3 GPa or 4.5 GPa, and 
finally disappears at a higher pressure of 6 GPa. 
This material is doped with holes with the total doping level of 
$\delta=0.25$, where $\delta$ is defined as the deviation of the 
density of electrons from the half-filling.
It has been proposed that at ambient pressure the holes are mostly in the 
chains, while high pressures cause the carrier to transfer 
into the ladders\cite{Kato}. 
If this is the case, and if most of the holes are 
transferred to the ladders at 6 GPa,
the experimental result is consistent with the present picture, since 
there is no enhancement of the 
pairing correlation for $\delta=0$ and $\delta\sim0.3$ due to the Umklapp 
processes as we have seen.
Evidently, further investigation especially in the large-$U$ regime
is needed to justify this speculation.

\section{Weak-coupling study of correlation functions 
in the three-leg Hubbard ladder}

In this section, we study correlation functions 
for the three(odd)-leg Hubbard ladder by the weak-coupling theory. 
As discussed in Sec.I, 
an increasing fascination in ladder systems 
has been caused by an `even-odd' conjecture 
for the existence of spin gap by Schulz\cite{SchulzAF} 
and independently by Rice {\it et al.}\cite{Rice} at half-filling.
When the system is doped with carriers, it is naively supposed that 
an even-numbered ladder should exhibit 
superconductivity with the interchain singlet pairing as expected from 
the persistent spin gap, while an odd-numbered ladder 
should have the usual $2k_F$ SDW reflecting 
the gapless spin excitations.

Theoretically, however, whether the 
`even-odd' conjecture for superconductivity 
continues to be valid for triple chains 
remains an open question. There had been no results for the
pairing correlation function in the three-leg $t-J$ 
or Hubbard ladder (Fig.3).

On the other hand, Arrigoni has looked into a three-leg 
with weak Hubbard-type interactions by 
the usual perturbational renormalization-group technique,
which is quite similar to that developed
by Balents and Fisher for the two-leg case,\cite{Balents}
to conclude that gapless and gapful spin excitations 
{\it coexist} there.\cite{Arrigoni}

Namely, he has actually enumerated  the numbers of gapless 
charge and spin modes on the phase diagram spanned by 
the doping level and the interchain hopping.
He found that, at half-filling, one gapless spin mode exists for 
the interchain hopping comparable with the intrachain hopping, 
in agreement with some experimental results and theoretical expectations
(Fig.12). 
Away from the half-filling, on the other hand, 
one gapless spin mode is found to remain at the fixed point 
in the region where the fermi level intersects all the 
three bands in the noninteracting case. 
From this, Arrigoni argues that 
the $2k_F$ SDW correlation should decay as a power law
as expected from experiments.

\begin{figure}
  \begin{center}
    \leavevmode
\leavevmode\epsfysize=55mm \epsfbox{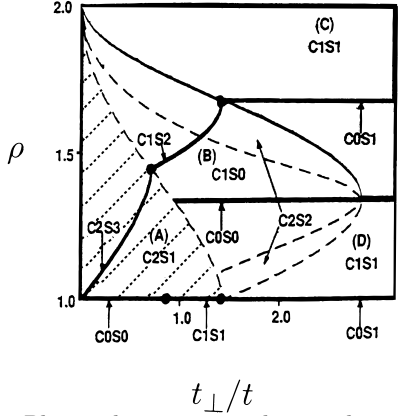}
     \caption{Phase diagram 
in the weak-coupling limit ($U\rightarrow0$) 
given in 
ref. [53]; 
the numbers of the gapless 
charge and spin modes ($x$ and $y$, respectively) are denoted as C$x$S$y$
and $\rho\equiv 2-n$ ($n$ is the band filling). 
In the dark region all the three bands cross the Fermi level.}
  \end{center}
  \label{fig:arri}
\end{figure}

On the other hand, 
his result also indicates that two gapful spin modes 
exist in addition. 
While a spin gap certainly 
favors a singlet superconducting (SS) correlation when there is 
only one spin mode, 
we are in fact faced here with an intriguing problem of what happens 
when gapless and gapful spin modes coexist, 
since it may well be possible that the presence of gap(s) in 
{\it some} out of multiple spin modes may be sufficient 
for the dominance of a pairing correlation.
Furthermore, as discussed below, the gaps of two spin modes emerge
as an  effect of the {\it pair-tunneling process} across 
the top and the bottom bands (Fig.4).
This is reminiscent of the two-leg case and of the Suhl-Kondo mechanism.
\cite{Suhl,Kondo}
These have motivated us, in this section, to actually 
look at the correlation functions 
using the bosonization method\cite{Review}
at the fixed point away from half-filling. 
Although in the three-leg case, we can consider the two boundary conditions 
across the legs, i.e., open boundary condition (OBC) 
and periodic boundary condition (PBC), 
here we concentrate on the open boundary condition 
(OBC) across the chains, where the central chain is 
inequivalent to the two edge chains. 
The reason is that we would like to (i) study the 
realistic boundary condition corresponds to cuprates, 
and (ii) to avoid the frustration 
introduced in the periodic three-legs.

We find that the interchain SS pairing 
across the central and edge chains is the dominant 
correlation, which is indeed realized due to 
the presence of the two gapful spin modes. 
On the other hand, the SDW correlation, 
which has a slowly-decaying power law for the intra-edge chain 
reflecting the gapless spin mode, coexists but is only subdominant.
\cite{Takashi1} 
Recently Schulz\cite{Schulz3} has independently
shown similar results for a subdominant $2k_F$ SDW and 
the interchain pairing correlations 
which are given in this section.

\subsection{Model and the Calculation}

The three-leg Hubbard model with OBC is defined by the following Hamiltonian,
\begin{eqnarray}
H=&&-t\sum_{\mu i\sigma}
(c^{\dagger}_{\mu i \sigma}c_{\mu i+1 \sigma}+{\rm h.c.})
\nonumber\\
&&-t_{\perp}\sum_{i\sigma}
(c^{\dagger}_{\alpha i\sigma}c_{\beta i\sigma}+
c^{\dagger}_{\beta i\sigma}c_{\gamma i\sigma}+{\rm h.c.})
\nonumber\\
&&+U\sum_{\mu i}n_{\mu i\uparrow}n_{\mu i\downarrow}, 
\end{eqnarray}
where $t (t_{\perp})$ is the intra-(inter-)chain hopping, 
$i$ labels the rung while $\mu = \alpha , \beta , \gamma$ labels the 
leg (with $\beta$ being the central one).  
In the momentum space we have
\begin{eqnarray}
H=&& \sum_{k\sigma}\left( -2t{\rm cos}(k)-\sqrt{2}t_{\perp}\right)
a^{\dagger}_{1k\sigma} a_{1k\sigma}\nonumber\\
&&-2t\sum_{k\sigma} {\rm cos}(k) a^{\dagger}_{2k\sigma} a_{2k\sigma}
\nonumber\\
&&+\sum_{k\sigma} \left( -2t{\rm cos}(k)+\sqrt{2}t_{\perp}\right)
a^{\dagger}_{3k\sigma}a_{3k\sigma}
\nonumber\\
&&+U\sum({\rm terms\  of\  the\  form}\ a^{\dagger}a^{\dagger}aa). 
\end{eqnarray}
Here $a_{jk\sigma}$ annihilates 
an electron with lattice momentum $k$ in the $j$-th band
($j=1,2,3$), where $a_{jk\sigma}$ is related to 
$c_{\mu k\sigma}$ (the Fourier transform of $c_{\mu i\sigma}$) 
through a linear transformation,
\begin{eqnarray}
\left( \begin{array}{c} 
c_{\alpha k\sigma} \\ c_{\beta k\sigma} \\ c_{\gamma k\sigma}
\end{array} \right)
=
\left( \begin{array}{ccc}
\frac{1}{2} & \frac{1}{\sqrt{2}} & \frac{1}{2}\\
\frac{1}{\sqrt{2}} & 0 & -\frac{1}{\sqrt{2}}\\
\frac{1}{2} & -\frac{1}{\sqrt{2}} & \frac{1}{2}
\end{array} \right)
\left( \begin{array}{c}
a_{1k\sigma} \\ a_{2k\sigma} \\ a_{3k\sigma}
\end{array} \right). \label{33}
\end{eqnarray}
Hereafter we linearize the band structure 
around the fermi points as usual and neglect 
the difference in the fermi velocities of three bands, 
as is done for calculating the correlation functions directly 
in the weak-coupling theory 
for the two-leg case,\cite{Fabrizio,Schulz2}
which will be acceptable for the weak interchain hopping. 
These approximations enable us to  
calculate the correlation functions. 
The difference in fermi velocities of three bands will 
not be important qualitatively as long as we consider
the case where three bands cross the fermi energy, 
for which Arrigoni's result falls on the same strong coupling fixed point 
on the plane of interchain hopping and filling.  
In the following, we focus on the case 
in which all of three bands are away from half-filling. 

The part of the Hamiltonian, $H_{\rm d}$, that can be 
diagonalized in the bosonization only includes 
forward-scattering processes in the band picture, 
and has the form
\begin{eqnarray}
H_{\rm d}&=&H_{\rm spin}+H_{\rm charge}, \nonumber\\
H_{\rm spin}&=&\sum_i\frac{v_{\sigma i}}{4\pi}\int\ dx
[\frac{1}{K_{\sigma i}}(\partial_x\phi_{i+})^2+
K_{\sigma i}(\partial_x\phi_{i-})^2], \\
H_{\rm charge}&=&\sum_i\frac{v_{\rho i}}{4\pi}\int\ dx
[\frac{1}{K_{\rho i}}(\partial_x\chi_{i+})^2
+K_{\rho i}(\partial_x\chi_{i-})^2].\nonumber
\end{eqnarray}
Here $\phi_{i+}$ is the spin phase field 
of the $i$-th band, $\chi_{i+}$ is the diagonal charge phase field, 
while 
$\phi_{i-}$($\chi_{i-}$) is the field dual to $\phi_{i+}$($\chi_{i+}$), 
$K_{\sigma i}$($K_{\rho i}$) the correlation exponent 
for the $\phi$($\chi_i$) phase 
with $v_{\sigma i}$($v_{\rho i}$) being their velocities. 
For the Hubbard-type interaction, we 
have $v_{\sigma i}=v_F$, $K_{\sigma i}$=1 for all $i$'s, while 
$v_{\rho 1}=v_F$, $v_{\rho 2}=v_F\sqrt{1-4g^2}$, 
$v_{\rho 3}=v_F\sqrt{1-g^2/4}$, $K_{\rho 1}=1$, 
$K_{\rho 2}=\sqrt{(1-2g)/(1+2g)}$,
$K_{\rho 3}=\sqrt{(1-g/2)/(1+g/2)}$, where $g=U/2\pi v_F$ 
is the dimensionless coupling constant.
The derivation of the above equation is given in the Appendix.

The diagonalized charge field $\chi_{i\pm}$ is linearly 
related to the initial charge field $\theta_{i\pm}$
of the $i$-th band as
\begin{eqnarray}
\left( \begin{array}{c}
\theta_{1\pm}\\ \theta_{2\pm}\\ \theta_{3\pm}
\end{array} \right)
=
\left( \begin{array}{ccc}
\frac{1}{\sqrt{2}} & \frac{1}{\sqrt{3}} & \frac{1}{\sqrt{6}}\\
                 0 & \frac{1}{\sqrt{3}} & -\sqrt{\frac{2}{3}}\\
-\frac{1}{\sqrt{2}} & \frac{1}{\sqrt{3}} & \frac{1}{\sqrt{6}}
\end{array} \right)
\left( \begin{array}{c}
\chi_{1\pm}\\ \chi_{2\pm}\\ \chi_{3\pm}
\end{array} \right),\label{35}
\end{eqnarray}
where both $\theta_{i\pm}$ and $\phi_{\pm}$ are related to the field 
operator for electrons $\psi_{i+(-)\sigma}$, which annihilates an 
electron on the right-(left-) going branch in band $i$ as 
\begin{eqnarray}
\psi_{i+(-)\sigma}(x)&=&\frac{\eta_{i+(-)\sigma}}{2 \pi \Lambda}
{\rm exp}\{ \pm ik_{iF}x\nonumber\\ &&\pm \frac{i}{2}
[ \theta_{i+}(x) \pm \theta_{i-}(x)+\sigma (\phi_{i+}(x)\pm\phi_{i-}(x))]\}.
\end{eqnarray}
Here the $\eta_{ir\sigma}$'s are Haldane's $U$ operators
\cite{Hal1D1}
which ensure the anti-commutation relations between electron 
operators through the relation, 
$\{\eta_{ir\sigma},\eta_{i'r'\sigma'}\}_+=
2\delta_{ii'}\delta_{rr'}\delta_{\sigma\sigma'}$, 
$\eta_{ir\sigma}^{\dagger}=\eta_{ir\sigma}$.

There are still many scattering processes
corresponding to both the backward and the pair-tunneling scattering 
processes, which cannot be treated exactly. 
Arrigoni examined the effect of such scattering 
processes by the perturbational renormalization-group technique.
He found that the backward-scattering interaction within 
the first or the third band turn from positive to negative 
as the renormalization is performed 
and that the pair-tunneling processes across the first 
and third bands also become relevant. 
As far as the relevant scattering processes are concerned, 
the first (third) band plays the role of the bonding 
(anti-bonding) band in the two-leg case. 
At the fixed point the Hamiltonian density, $H^{*}$, 
then takes the form, in term of the phase variables, 
\begin{eqnarray}
H^{*}&=&-\frac{g_{b}(1)}{\pi^2 \Lambda^2}{\rm cos}(2\phi_{1+}(x))
            -\frac{g_{b}(3)}{\pi^2 \Lambda^2}{\rm cos}(2\phi_{3+}(x))
\nonumber\\
          &&+\frac{2g_{ft}(1,3)}{\pi^2 \Lambda^2}
            {\rm cos}(\sqrt{2}\chi_{1-}(x))
            {\rm sin}\phi_{1+}(x){\rm sin}\phi_{3+}(x),
\end{eqnarray}
where both $g_{b}(1)$ and $g_{b}(3)$ are negative large quantities, 
and $g_{ft}(1,3)$ is a positive large quantity. 

This indicates that the phase fields 
$\phi_{1+}$, $\phi_{3+}$, and $\chi_{1-}$ are
long-range ordered and fixed 
at $\pi/2$, $\pi/2$, and $\pi/\sqrt{2}$, respectively, 
which in turn implies that 
the correlation functions that contain 
$\phi_{1-}$, $\phi_{3-}$, and $\chi_{1+}$ fields 
decay exponentially. The renormalization procedure will 
affect the velocities and the critical exponents 
for the gapless fields, $\chi_{2\pm}$, $\chi_{3\pm}$, and $\phi_{2\pm}$, 
so that we should end up with renormalized 
$v^*$'s and $K^*$'s. 

In principle, the numerical values of renormalized $v^*$'s and $K^*$'s  
for finite $g$ may be obtained 
from the renormalization equations 
as has been attempted for a double chain by Balents and Fisher 
\cite{Balents}, although it would be difficult in practice.
However, at least in the weak-coupling limit, $g\rightarrow 0$, 
to which our treatment is meant to fall upon, 
we shall certainly have $v^{*}\simeq v_F$ and $K^* \simeq 1$ 
for {\it gapless} modes even after the renormalization procedure.

\subsection{Results for the Correlation Functions}

Now we are in position to calculate the correlation functions,
since the gapless fields have 
already been diagonalized, while the remaining gapful fields have 
the respective expectation values. 
The details of the calculation of the correlation functions are
given in the Appendix.
The two-particle correlation functions which include the following
two particle operators in the band description are shown to
have a power-law decay:

\noindent (1) operators constructed from two operators involving 
only the second band
 (since the charge and the spin phases are both gapless, electrons 
in this band should have the usual TL-liquid behavior),

\noindent (2) order parameters of singlet superconductivity 
within the first or 
third band, $\psi_{1+\uparrow(\downarrow)}\psi_{1-\downarrow(\uparrow)}$,
$\psi_{3+\uparrow(\downarrow)}\psi_{3-\downarrow(\uparrow)}$. 

\noindent As a result, the order parameters that possess power-law decays 
should be the following, 

\noindent (A) The correlations within each of the two edge 
($\alpha$ and $\gamma$) chains or across the two edge chains: 

\noindent (a) $2k_F$ CDW, \par
$O_{{\rm intra} 2k_F {\rm CDW}}=\psi_{\alpha(\gamma)+\uparrow}^{\dagger}
               \psi_{\alpha(\gamma)-\uparrow}$;\par
$O_{\rm inter CDW}=\psi_{\alpha(\gamma)+\uparrow}^{\dagger}
               \psi_{\gamma(\alpha)-\uparrow},$\par
\noindent (b) $2k_F$ SDW, \par
$O_{\rm intra SDW}=\psi_{\alpha(\gamma)+\uparrow}^{\dagger}
               \psi_{\alpha(\gamma)-\downarrow}$;\par 
$O_{\rm inter SDW}=\psi_{\alpha(\gamma)+\uparrow}^{\dagger}
               \psi_{\gamma(\alpha)-\downarrow}$,\par
\noindent (c) singlet pairing (SS), \par
$O_{\rm intra SS}=\psi_{\alpha(\gamma)+\uparrow}
               \psi_{\alpha(\gamma)-\downarrow}$;\par
$O_{\rm inter SS}=\psi_{\alpha(\gamma)+\uparrow}
               \psi_{\gamma(\alpha)-\downarrow}$,\par
\noindent (d) triplet pairing (TS), \par
$O_{\rm intra TS}=\psi_{\alpha(\gamma)+\uparrow}
               \psi_{\alpha(\gamma)-\uparrow}$;\par
$O_{\rm inter TS}=\psi_{\alpha(\gamma)+\uparrow}
               \psi_{\gamma(\alpha)-\uparrow}$,\par

\noindent (B) The $4k_F$ CDW which is written with 
four electron operators, \par
$O_{4k_F{\rm CDW}}=
\psi_{\nu+\uparrow}^{\dagger}\psi_{\nu+\downarrow}^{\dagger}
\psi_{\nu-\uparrow}\psi_{\nu-\downarrow}\ \ \  (\nu=\alpha,\beta,\gamma)$, \par

\noindent (C) The singlet pairing across 
the {\it central} chain ($\beta$) and edge chains
(Fig.13), \par
$O_{\rm CESS}=\sum_{\sigma}\sigma(\psi_{\alpha+\sigma}+\psi_{\gamma+\sigma})
\psi_{\beta-,-\sigma}$. 

\begin{figure}
  \begin{center}
\leavevmode\epsfysize=30mm \epsfbox{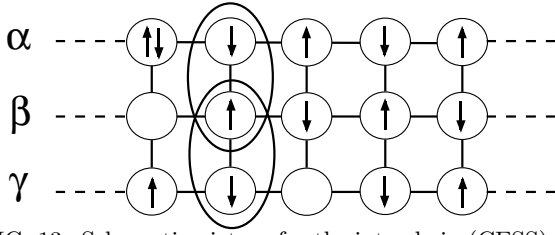}
     \caption{Schematic picture for the interchain 
(CESS) pairing in the doped three-leg 
Hubbard ladder.}
  \end{center}
  \label{fig:pairing3}
\end{figure}

In the band picture we can rewright 
$O_{\rm CESS}$ as comprising 
\begin{eqnarray}
O_{\rm CESS} \sim \sum_{\sigma}\sigma(\psi_{1+\sigma}\psi_{1-,-\sigma}
-\psi_{3+\sigma}\psi_{3-,-\sigma}). 
\end{eqnarray} 
We cannot easily name the symmetry of the pairing, 
although we naively might call this pairing d-wave-like 
in a similar sense as in the two-leg case, 
in which a pair is called d-wave 
when the pairing, in addition to being off-site, consists of 
a bonding band and an anti-bonding band pairs 
with opposite signs.\cite{Balents,Fabrizio,Nagaosa,Schulz2,Noack3}
Thus the edge-chain SDW correlation has a power-law decay, 
while the SDW correlation within the central chain 
decays exponentially since it consists of 
the terms containing $\phi_{1-}$ and/or $\phi_{3-}$ phases.  
Although we consider the case away from half-filling, 
the SDW correlation should obviously be 
more enhanced at half-filling. 
The experiments at half-filling do not contradict the present results,
since the experiments should detect the total 
SDW correlation of all the chains and the 
SDW correlation is more enhanced at half-filling. 
However the present theory corresponds only to the 
infinitesimally small interaction in principle, 
although the actual cuprates have a strong interactions between electrons. 

Intra- or inter-edge correlation functions 
have to involve forms bilinear in $a_{2k\sigma}$ in eq.(\ref{33}).  
They are described in terms of the charge field 
$\theta_2$ for the second band, which does not contain $\chi_1$, 
a phase-fixed field (see eq.(\ref{35})).  
Thus the edge-channel correlations are 
completely determined by the character of the second band 
(the Luttinger-liquid band), while the other phase fields, being gapful, 
are irrelevant. 
The final result for the edge-channel correlations 
at large distances, up to $2k_F$ oscillations, is 
as follows regardless of whether the correlation is intra- or inter-edge: 
\begin{eqnarray}
\langle O_{2k_F{\rm CDW}}(x)O_{2k_F{\rm CDW}}^{\dagger}(0)\rangle&\sim& 
x^{-\frac{1}{3}(K_{\rho 2}^*+2K_{\rho 3}^*)-1}, 
\nonumber\\ 
\langle O_{\rm SDW}(x)O_{\rm SDW}^{\dagger}(0)\rangle&\sim& 
x^{-\frac{1}{3}(K_{\rho 2}^*+2K_{\rho 3}^*)-1}, 
\nonumber\\ 
\langle O_{\rm SS}(x)O_{\rm SS}^{\dagger}(0)\rangle&\sim& 
x^{-\frac{1}{3}(\frac{1}{K_{\rho 2}^*}+\frac{2}{K_{\rho 3}^*}) 
-1}, \\
\langle O_{\rm TS}(x)O_{\rm TS}^{\dagger}(0)\rangle&\sim&
x^{-\frac{1}{3}(\frac{1}{K_{\rho 2}^*}+\frac{2}{K_{\rho 3}^*})
-1}. 
\nonumber
\end{eqnarray}
(where we have put $K_{\sigma}^*=1$ for the present 
spin-independent interaction.\cite{Schulz3})
In addition, the $4k_F$ CDW correlation decays as 
\begin{eqnarray}
\langle O_{4k_F{\rm CDW}}(x)O_{4k_F{\rm CDW}}^{\dagger}(0)\rangle\sim
x^{-\frac{2}{3}(2K_{\rho 2}^*+K_{\rho 3}^*)}.
\end{eqnarray}
By contrast, if we look at the pairing $O_{\rm CESS}(x)$ across the 
central chain and the edge chains, 
this pairing, which circumvents the on-site repulsion 
and is linked by the resonating valence bonding across 
the neighboring chains, 
is expected to be stronger than other correlations as in the two-leg case. 
The correlation function for $O_{\rm CESS}(x)$ 
is indeed calculated to be 
\begin{eqnarray}
\langle O_{\rm CESS}(x)O_{\rm CESS}^{\dagger}(0) \rangle
\sim x^{-\frac{1}{3}(\frac{1}{K_{\rho 2}^*}+\frac{1}{2K_{\rho 3}^*})}.
\end{eqnarray}
From the calculations given in the Appendix, we can see 
that the interchain pairing exploits the charge gap 
and the spin gaps to reduce the exponent of the correlation
function, in contrast to the intra-leg pairing. 
In addition to that, we also find that the roles of the first 
(third) band corresponds to those of the bonding (anti-bonding) band 
in the two-leg case, as far as the dominant pairing correlation is 
concerned.
If we consider the weak-interaction limit ($U\rightarrow +0$)
as in the two-leg case, all the 
$K^*$'s will tend to unity, where the CESS 
correlation decays as $x^{-1/2}$ while those of other correlations 
decays as $x^{-2}$ at long distances.
Thus, at least in this limit, the CESS correlation dominates over
the others. 
The duality (which dictates that the pairing and density-wave 
exponents are reciprocal of each other\cite{Nagaosa}) 
is similar to that in the two-chain case, in which the 
interchain SS decays as $x^{-1/2}$ while that of the $4k_F$ CDW 
decays as $x^{-2}$.


In this section, we have studied correlation functions 
using the bosonization method at the renormalization-group 
fixed point, which was obtained by Arrigoni, 
away from half-filling in the region where the
fermi level intersects all the three bands in the non-interacting case. 
We found that the interchain singlet pairing 
{\it across} the central chain and either of the edge chains 
is the dominant correlation contrary to the naive 
`even-odd' conjecture for the superconductivity in ladder systems, 
while the SDW correlations in two edge chains coexist but are subdominant. 
The power law decay of the SDW correlation does not contradict with 
the even-odd conjecture at the half-filling, where 
the Umklapp scattering play a important role resulting in 
an enhancement of the SDW correlation.

The renormalization study is valid only 
for infinitesimally small interaction 
strengths and sufficiently small interchain hoppings 
in principle, while the actual cuprates 
have strong interactions between electrons, 
so that the relevance of 
the present results to the real materials is uncertain. 
However the present study suggests 
an important theoretical message that the dominance of 
superconductivity only requires 
the existence of gap(s) in {\it some} spin modes 
when there are multiple modes in multi-leg ladder systems 
no matter whether the number of legs is odd or even.

\section{QMC study of the pairing 
correlation in the three-leg Hubbard ladder}

In the previous section, we have discussed the correlation 
functions in the three-leg Hubbard ladder within 
the weak-coupling theory. A key point in the previous section is 
that the gapless and the gapful spin excitations coexist in a three-leg ladder 
and the modes give rise to a peculiar situation 
where a specific singlet pairing 
{\it across} the central and edge chains (CESS pairing), 
that may be roughly a d-wave pairing, 
is dominant, while the $2k_F$ SDW on the edge chains simultaneously 
shows a subdominant but still long-tailed (power-law) decay 
associated with the gapless spin mode. 
This result is stimulating since it 
serves as a counter-example of a naive `even-odd' conjecture.

However, there is a serious question about 
these weak-coupling results as 
discussed in the two-leg case in Sec.II. 
First, only for infinitesimally small interactions and 
sufficiently small hoppings are the results
in Sec.III guaranteed to be valid in principle. 
Furthermore, when there is a gap in the excitation, 
the renormalization flows into a strong-coupling regime, 
so that the weak-coupling 
theory might break down even for small $U$.   
Hence it is imperative to study the problem from an independent 
numerical method for an intermediate strength of the Hubbard $U\sim t$ 
and an interchain hopping $t_{\perp}\sim t$. 
Although such a comparison of the numerical result for $U\sim t$  
with the weak-coupling theory has been done for 
the two-leg system in Sec.II, this does not necessarily 
shed light on the situation in the three-leg case, where 
gapless and gapful modes coexist. 
In this section, it is shown that the QMC 
result for the three-leg Hubbard ladder
indeed turns out to exhibit an enhancement of 
the pairing correlation even for finite coupling constants, 
$U/t=1\sim 2$.\cite{Takashi2} 

In addition, we also study the effects of various Umklapp 
processes at special fillings as in the two-leg case 
keeping in mind the above Arrigoni's work\cite{Arrigoni} which also 
studied the effects of some Umklapp processes with the 
weak-coupling theory. 

Throughout this section, 
we concentrate on the case in which
all three bands cross the Fermi surface 
to explore the properties of a three-band system.

The projector Monte Carlo method is employed
\cite{Hirsch,MonteCarlo}
to investigate the ground-state
pairing correlation function 
$P(r)\equiv\langle O^{\dagger}_jO_{j+r}\rangle$, 
where $O_i\equiv O_{CESS}(i)
=(c_{\alpha i\sigma}+c_{\gamma i\sigma})c_{\beta i-\sigma}- 
(c_{\alpha i-\sigma}+c_{\gamma i-\sigma})c_{\beta i\sigma}$. 
We assume the periodic boundary condition
along the chain direction, $c_{N+1}\equiv c_1$, where $N$
is the number of rungs. We first consider the case where the 
intra- and the inter-band Umklapp processes are irrelevant because
that is  where the above-mentioned weak-coupling 
theory is valid. 
The details of the QMC calculation are similar
to those for our QMC study for the two-leg case.
Specifically, the negative-sign problem makes the QMC calculation 
feasible for $U\leq 2t$ as in the two-leg case.  
We set $t=1$ hereafter.

In the two-leg case with a finite $U$, 
we have found an interesting property for 
finite systems: 
the pairing correlation is enhanced 
in agreement with the weak-coupling theory 
only when the single-electron energy
levels of the bonding and the anti-bonding bands lie close to 
each other around the Fermi level 
(which is certainly the case with an infinite system).
When the levels are misaligned 
(for which a 5\% change in $t_{\perp}$ is enough), 
the enhancement of the pairing correlation dramatically vanishes.  
In the weak-coupling theory, the ratio of 
the spin gap to the level offset is assumed to be infinitely large 
at the fixed point of the renormalization flow, 
so that the spin gap should naturally be detectable in finite systems
only when the level offset is smaller than the gap.

We have found that this applies to the three-leg ladder 
as well, i.e., 
the pairing correlation is enhanced 
when the single-electron levels of 
the first and third bands lie close to each other. 
Hence we concentrate on such cases hereafter.

In the beginning we show in Fig.14
the result for $P(r)$ for $t_{\perp}=0.92$ with $U=1$ with the band filling
$n=0.843=86$ electrons/(34 rungs$\times$ 3 sites). 
For this choice of $t_{\perp}$ the levels in the
first and the third bands lie close to each
other around the Fermi level within 0.01.  
We can see that a large enhancement over 
the $U = 0$ result at large distances indeed exists. 
This is the key result of this section.

\begin{figure}
  \begin{center}
\leavevmode\epsfysize=70mm \epsfbox{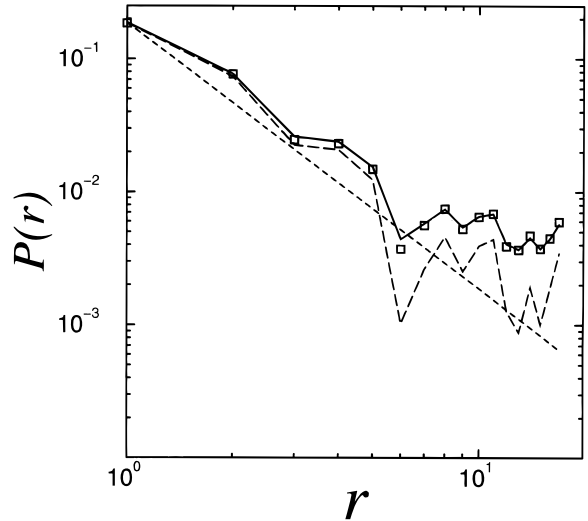}
      \caption{QMC result for the pairing correlation function, 
$P(r)(\Box)$, plotted against the
real space distance $r$ in a three-leg Hubbard ladder 
with 34 rungs having 86 electrons for $U = 1$
with $t_{\perp} = 0.92$. 
The dashed line is the noninteracting result 
for the same system size, while the
straight dashed line represents $\sim r^{-2}$. 
The solid line is a fit to a trial function (see text).
}
  \end{center}
  \label{fig:fig41}
\end{figure}

Although it is difficult
to determine the decay exponent of $P(r)$, 
we can fit the data by supposing a trial function 
as expected from the weak-coupling theory 
as we did in the two-leg case,
\begin{eqnarray}
P(r)&=&\frac{1}{\pi^2}\sum_{d=\pm}\{cr_d^{-1/2}\nonumber\\
&&+[(2-c)-\cos(2k_{F1} r_d)-\cos(2k_{F3} r_d)]r_d^{-2}\}
\label{tri2}
\end{eqnarray}

Here $k_{F1}(k_{F3})$ is the non-interacting Fermi wave number
of the first (third) band, while a constant $c$, 
which should vanish for $U=0$, 
is here least-square fit (by taking logarithm of the data)
as $c = 0.05$. 
As in the two-leg case, since we assume the periodic boundary condition, 
we have to consider contributions from both ways around, 
so there are two distances between the 0-th and 
the $r$-th rung, i.e.,
$r_+ = r$ and $r_- = N - r$. 
The overall decay should be $1/r^2$ 
as in the single-chain case, while the term 
$c/r^{1/2}$, the dominant 
correlation at large distances, is borrowed from 
the weak-coupling result.\cite{Takashi1,Schulz3} 
The QMC result for a finite $U = 1$ 
fits to the trial form (solid line in Fig.14) 
surprisingly accurately.  
A finite $U$ may give some corrections to these functional forms, 
but even when we best-fit the exponent itself as $c/r^{\alpha}$ 
in place of $c/r^{1/2}$, 
we obtain $\alpha <0.7$ with a similar accuracy.

In Fig.15, we show the result for a larger 
interaction $U = 2$. 
The result again shows an enhanced pairing correlation at large
distances.
However, the enhancement is slightly reduced 
than that in the $U = 1$ case. 
This is consistent with the weak-coupling theory, 
in which $K^*_{\rho}$'s should decrease with $U$.

\begin{figure}
 \begin{center}
\leavevmode\epsfysize=70mm \epsfbox{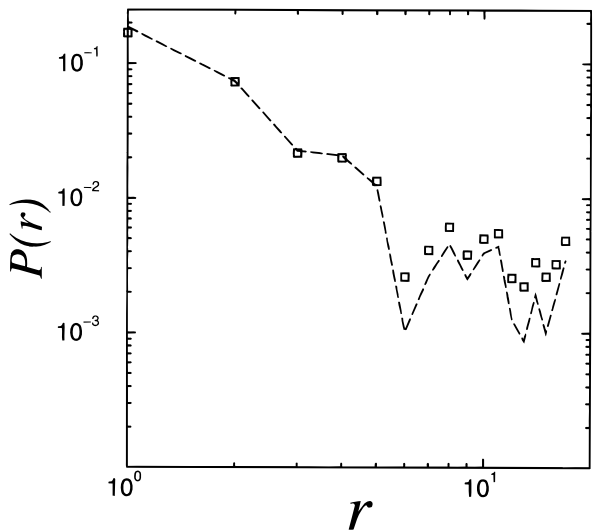}
    \caption{
Plot similar to that in Fig.14  for $U = 2$.
}
  \end{center}
\label{fig:fig42}
\end{figure}

\begin{figure}
 \begin{center}
\leavevmode\epsfysize=70mm \epsfbox{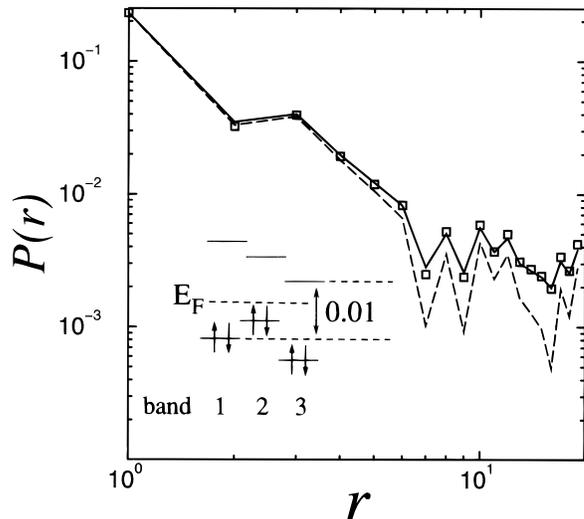}
     \caption{Similar plot as in Fig.14 for a 38 rungs system 
having 82 electrons for $U = 1$ with $t_{\perp} = 0.685$.
The inset schematically depicts the positions of 
energy levels for the noninteracting case.}
  \end{center}
\label{fig:fig43}
\end{figure}

Furthermore, we study if the presence of the second band around 
$E_F$ can be detrimental to superconductivity.  
In Fig.16, 
we make the single-electron energy levels of all the three bands 
lie close to each other around
the Fermi level. 
This is accomplished here 
for $t_{\perp}= 0.685$ 
and the band filling $n = 0.719 = 82$ electrons/(38 rungs $\times$
3 sites). 
The highest occupied level of the second band then lies 
between that of the first band 
and the lowest unoccupied level of the
third band (lying above the highest 
occupied level of the first band by as small as 0.01, 
inset of Fig.16).  

The result in Fig.16 for $U = 1$ shows that 
the pairing correlation is enhanced as well. Thus we may
consider that the second band does not hinder 
the enhancement of the pairing correlation
in other bands. This is also consistent with the weak-coupling theory, 
in which all of the scattering processes connected with 
the second band are irrelevant.   
The fit of the correlation function to the trial one is again 
excellent with $c = 0.03$.

In the remaining of this section, we discuss the effects of the Umklapp 
processes at special fillings to clarify the doping dependence.
In the three-leg Hubbard model, 
the Umklapp processes can play an important role at specific band fillings. 

Arrigoni\cite{Arrigoni} also studied the effect of Umklapp processes
within the weak-coupling theory, although he did not calculate 
the correlation functions directly (see Fig.12). 
He studied two cases that have the relevant Umklapp processes.

(i) the half-filled case.
 
(ii) the case when the bottom band is half-filled, 
in which the intraband Umklapp process may become relevant within 
the first band.

In both cases, the Umklapp processes become 
relevant. 
In the former case, the system  has a gapless 
spin excitation suggesting a power-law decaying AF correlation, 
as discussed by Arrigoni,
for $t_{\perp}\sim t$, a region of interest.
The phase is called a `C1S1' regime, since there is 
one gapless mode in charge or spin mode. 
Although the existence of a gapless charge mode
suggests that the system is not an insulator, a full charge 
gap is expected to appear for sufficiently large $U$. 
Then, the pairing correlation will be suppressed, 
although the direct calculation has not been done.
(As discussed in Sec.II, there are two kinds of charge gaps. 
The one, which is  produced by the pair-tunneling processes, 
favors the pairing correlation as in the previous section, 
while the other, which is produced by the Umklapp processes,
suppress the pairing correlation as in the present section.) 

In the latter case, the Umklapp process also becomes 
relevant resulting in a `C2S3' phase with the 
two gapless charge modes and three gapless spin modes
and the singlet-pair of holes or electrons is 
prevented from binding.
Thus the pairing correlation will be not 
enhanced because of the absence of gapful spin mode(s).
This situation is reminiscent of the two-leg case, in which 
the spin gap is destroyed when the bonding band is half-filled.
The Umklapp process within the first band in the present case 
is identified with that in the bonding band in the two-leg case, 
since the first (third) band corresponds to the bonding (anti-bonding) 
band as far as the pairing correlation is concerned as discussed in 
Sec.III.

Given this situation, our motivation here is to look at the pairing 
correlation and, in addition, to explore in which case the 
Umklapp scattering does or does not affect the pairing correlation.
We study the above two cases and also 
study the case in which the second band is half-filled. 
Namely, we wish to see whether 
both the first and the third bands, which involve the pairing 
order parameter, are affected by an indirect effect of the Umklapp process 
in the second band. 
Such a situation emerges in ladder systems with three or larger
number of legs. 

We have tuned the value of $t_{\perp}$ to ensure that 
the level offset $(\Delta\varepsilon)$ 
between the first and the third bands at the Fermi level 
is as small as O($0.01$) for $U=0$ to 
single out the effect of Umklapp processes from
those due to large values of $\Delta\varepsilon$.
Ideally, we should make the single electron energy 
levels of all the three bands lie close to each other around
the Fermi level but that is impossible within the 
tractable system sizes. 
However, when the Umklapp process is relevant within 
the first or the third band, 
the aligned levels of the bands should favor the pairing, so that 
if the pairing correlation is suppressed, we can infer that the suppression 
is not an artifact. On the other hand, when the level of the second 
band is 
misaligned from $E_F$, one may naively think that the effect of the Umklapp 
process within the band becomes obscured. 
However, the effect of the Umklapp processes should in fact be enhanced 
when the highest occupied level  derivate from $E_F$, since the Umklapp 
processes become well-defined if the highest occupied level is doubly occupied.

In the beginning we look at the half-filling (Fig.17). 
Indeed, no enhancement of the pairing 
correlation is found and the over all decaying form is 
similar to the $U=0$ result as in the two-leg case at half-filling.

\begin{figure}
 \begin{center}
\leavevmode\epsfysize=70mm \epsfbox{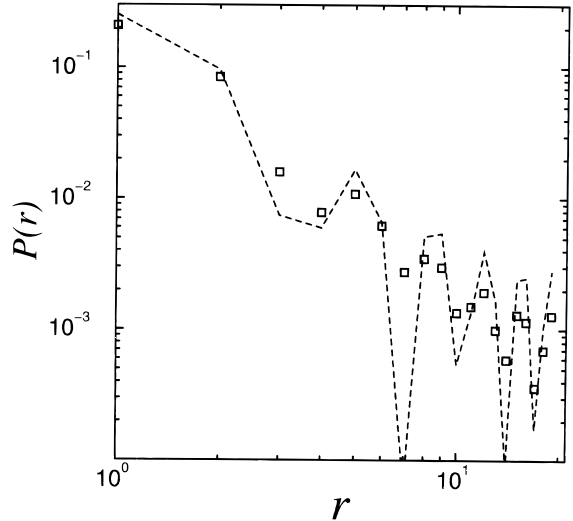}
     \caption{Pairing correlation $P(r)$ ($\Box$) against $r$ 
for a 38 rungs system for $U=2$ with $t_{\perp}=0.955$
and 114 electrons (half-filling). 
The dashed line represents the noninteracting result.
}
  \end{center}
\label{fig:fig44}
\end{figure}

When $n$ is decreased to make  
the second band half-filled, the enhanced 
pairing correlation is found (Fig.18). 
Possibilities are either 
the Umklapp process is not relevant,  
or it is relevant but does not affect 
the other bands. 
In the latter case, a density-wave correlation might be dominant 
due to a charge gap opening by the Umklapp scattering. 
However, at least in the sense of the weak-coupling theory, 
the charge gap in the second band only enhances
the density-wave correlation at long distances from $r^{-2}$ 
to $r^{-1}$ 
(unity, the value of the exponent, comes from the gapless spin mode 
and it should be independent to $U$)
in the weak-coupling limit and thus the pairing correlation 
may still remain dominant for small $U$.

\begin{figure}
 \begin{center}
\leavevmode\epsfysize=70mm \epsfbox{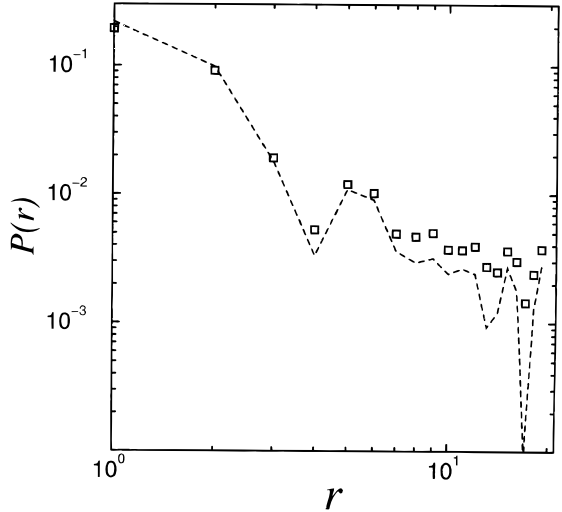}
      \caption{Pairing correlation $P(r)$ ($\Box$) against $r$ 
for a 38 rungs system for $U=2$ with $t_{\perp}=0.87$
and 110 electrons (the half-filled second band). 
The dashed line represents the noninteracting result.
}
  \end{center}
\label{fig:fig45}
\end{figure}

Although we cannot decide which of the above two possibilities applies, 
we do have a unique situation where the pairing correlation is enhanced 
{\em despite} the Umklapp processes being possible. 
This interesting situation does not appear in the two-leg ladder.

Lastly, we study the case so that the first band is half-filled 
when $n$ is further decreased and $t_\perp$ is also decreased
(Fig.19). 
In this case, we again observe no enhancement in $P(r)$,
as expected from the weak-coupling theory and the above discussion.

\begin{figure}
 \begin{center}
\leavevmode\epsfysize=70mm \epsfbox{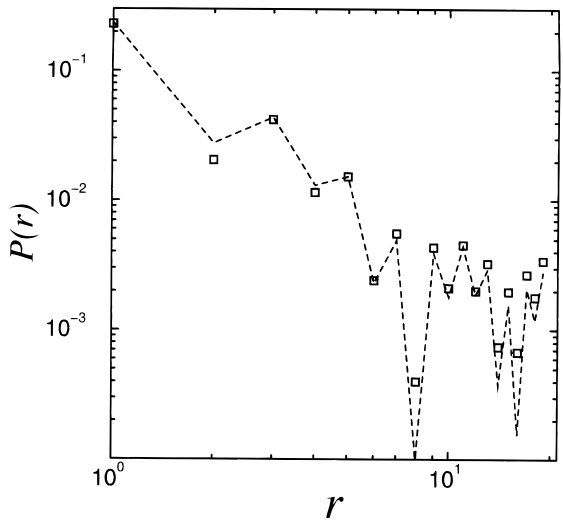}
      \caption{Pairing correlation $P(r)$ ($\Box$) against $r$ 
for a 38 rungs system for $U=2$ with $t_{\perp}=0.725$
and 74 electrons (the half-filled first band). 
The dashed line represents the noninteracting result.
}
  \end{center}
\label{fig:fig46}
\end{figure}

In this section, we have shown with the projector quantum Monte Carlo method
that the enhancement of the 
pairing correlation expected from the results in the 
previous section is indeed found even for the intermediate 
interaction strengths($U\sim 2t$) and the interchain 
hoppings($t_{\perp}\sim t$). 
The features of the enhancement is similar to that in the two-leg case. 

We have also studied the cases where the Umklapp processes 
can be relevant. 
Especially, we found that the enhancement of the pairing 
correlation is not affected by the intraband Umklapp process 
within the second band which does not involve the pairing order parameter.

\section{Pairing correlation in finite systems --- 
a case study for the 1D attractive Hubbard model}

As we have seen in sections 2 (two-leg ladder) and 4 (three-leg ladder), 
QMC result for the pairing 
correlation in Hubbard ladders are  
consistent with the weak-coupling prediction when 
the highest occupied level 
and the lowest unoccupied level (called the level offset 
$\Delta\epsilon$ hereafter) 
in the relevant free-electron bands are made to lie close to each other. 
A similar situation is also found to occur 
in a QMC study of the single chain with distant transfers 
($t-t^{\prime}-U$ model).\cite{Arita}  

We believe that small level offsets 
should be required to mimic the thermodynamic limit (bulk systems), where 
this quantity is infinitesimal. 
However, we have to quantify the criterion systematically.  
Otherwise, the results obtained in the previous section 
may be taken as an uncontrolled finite size effect. 
In this section we actually 
give a circumstantial evidence 
for justifying that we can indeed quantify this, with which 
we can tell that the results in the previous sections 
are meaningful.  

We start with asking ourselves the following question.  
Let us take a model (such as the attractive Hubbard model 
with an attractive interaction $U<0$) 
that is known to superconduct.  
For a bulk system we know that the pairing correlation should 
have an slowly decaying asymptotic behavior for arbitrary 
$U\neq0$, but 
in a numerical calculation such as QMC the 
tractable system size is limited, so that it is 
inconceivable that a slowly decaying asymptotic behavior is obtained 
even for, say, $|U|=0.01t$. 
So the question is what is the requirement for such an 
asymptotic behavior to be detectable in finite size systems.

We shall conclude in this section 
that the level offset as compared with the {\it spin gap} is 
the key ingredient.  
The criterion will also shed light on 
numerical calculations for systems 
which may possibly have small spin gaps.

As a case study, we take here the 1D attractive Hubbard model. 
This is because the model is 
exactly solvable with the Bethe Ansatz\cite{Bethe}, 
so that we know the exact form of the 
correlation function as well as the value of the spin gap for 
arbitrary values of 
parameters.  
We compare this with QMC results 
for the pairing correlation for system sizes exceeding one 
hundred sites.  We deliberately choose 
small attractions to look into the case where 
the spin gap is small. 

Let us first briefly recapitulate the exact result for the 1D attractive 
Hubbard model, whose Hamiltonian is given by 
\begin{equation}
{\cal H}=-t\sum_{i\sigma}
(c_{i\sigma}^\dagger c_{i+1 \sigma}+{\rm h.c.})+
U\sum_i n_{i\uparrow}n_{i\downarrow}\;\;\;\;(U<0)
\end{equation}
in standard notations. We consider the case where the numbers of up-spin
and down-spin electrons coincide. In this case, the spin gap is present
for all the values of $U<0$ and the band filling $n$. 
At half filling $(n=1)$, the
pairing correlation decays as $1/r$ with the real space
distance $r$, regardless of the value of $U$.  
Note that the power is reduced abruptly 
by unity from the power (=2) for $U=0$ as soon as an infinitesimal 
$U$ is switched on.  
This is an effect of the spin gap, which opens 
as soon as an infinitesimal 
$U$ is switched on, this is due to
the fact that the spin phase is locked to give a long range order for
its contribution to the pairing correlation function 
for $U < 0$.  
For a general value of $n$ the pairing correlation decays as 
$1/r^{1/K_\rho}$, where $(1\leq)K_\rho(<2)$, an exponent appearing 
in the Tomonaga-Luttinger theory, 
monotonically increases with $|U|$, 
and decreases with $n(\leq 1)$.

We have adopted the 
projector Monte Carlo method to calculate the on-site singlet
pairing correlation function
\begin{equation}
P(r)=\langle c_{i\uparrow}^\dagger c_{i\downarrow}^\dagger
c_{i+r\downarrow} c_{i+r\uparrow}\rangle
\end{equation}
in the ground state. The calculation is free from the negative sign problem
because we are considering the case of attractive $U$.\cite{Hirsch}
All the calculations are performed for a 114-site system.
We consider the case where the closed shell condition 
(non-degenerate free-electron Fermi sea) is full-filled, 
so that there is a finite gap $\Delta\varepsilon$ between the 
highest occupied level and the lowest unoccupied level.
To evaluate the decaying power of the pairing correlation, 
we least-square fit the 
numerical result to $1/r^{\alpha}$ in the range $r=6\sim 27$ 
by taking logarithm of the data. 
In this range the periodic boundary condition
is seen to have little effect.  
Although a cosine-like fluctuation 
is obviously present in the 
data, fitting all of the points in a finite range should average
this effect out, namely, fitting $\cos(kr)/r^\beta$ to a form $a/r^\gamma$
would give $a\sim 0$.

In Fig.20, we first present results for half filling $(n=1)$, 
where the 
pairing correlation has an asymptotic behavior
like $\sim 1/r$ {\it regardless of} $U$.
Our QMC result for $U=-t$ shows that the power is 
$\alpha\sim 1.4$ in contradiction with 
the exact result. Only when $U$ becomes as large as 2 do we 
recover $\alpha\sim 1$.

\begin{figure}
  \begin{center}
\leavevmode\epsfysize=100mm \epsfbox{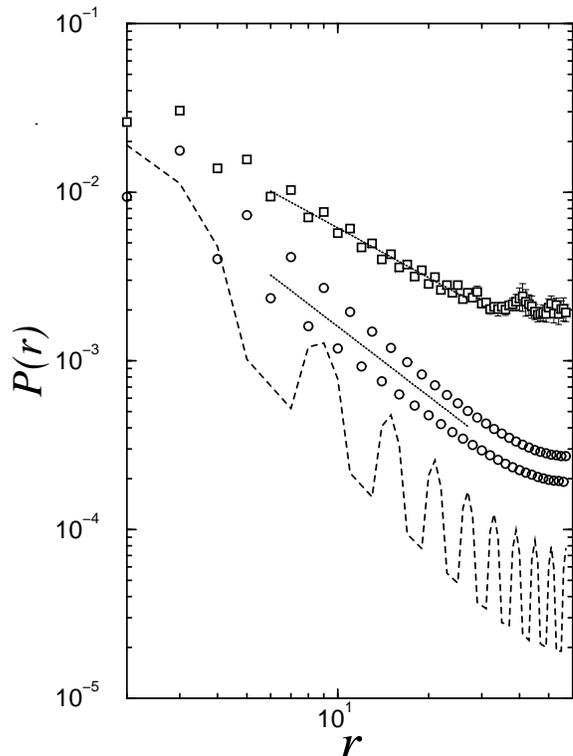}
\caption{
QMC result for
the on-site pairing correlation function, $P(r)$, plotted 
against the real space distance $r$ in a 114-site 
1D attractive Hubbard model at half filling. $U=0$ (dashed line),
$U=-t$ ($\bigcirc$), $U=-2t$ ($\Box$). Straight dashed lines are 
least squares fit with $1/r^\alpha$ for $r=6\sim 27$.
}
  \end{center}
\label{fig1}
\end{figure}

Now, the spin gap $\Delta_S$ for this system size at 
$U=-t (-2t)$  
estimated from the Bethe Ansatz is 0.12t (0.23t).
If we take the ratio of these values to the
level offset $\Delta\varepsilon$ around the Fermi level, 
which is $0.11t$, they give values 1.1 and 2.1, respectively.
This suggests that the pairing correlation, which is governed by 
the spin gap, cannot be clearly detected unless 
$\Delta_S/\Delta\varepsilon$ is sufficiently large, in
agreement with our speculation mentioned at the beginning of the 
present section.  In fact 
$\Delta_S/\Delta\varepsilon$ is infinite in a bulk system, 
and we also assume an infinite $\Delta_S/\Delta\varepsilon$ 
in the Tomonaga-Luttinger theory as well.

In Fig.21, in order to confirm the point away from the half filling, 
we have also looked at $n=1/3$,
where $\Delta\varepsilon(=0.055t)$ is smaller than that for $n=1$.
The power for the 
pairing correlation is now $\alpha\sim 0.85$ for $U=-t$ 
form the Bethe Ansatz.  
The QMC result for $U=-t$ gives $\alpha\sim 0.9$,
indicating that the numerical result
is rather reliable already at $U=-t$ in this case.  
If we look at the level offset, 
$\Delta_S/\Delta\varepsilon \sim 1.6$ is indeed greater than that for
$n=1$ and is closer to the case of $U=-2t$ for $n=1$.
This result further confirms our view.

\begin{figure}
  \begin{center}
\leavevmode\epsfysize=100mm \epsfbox{figure21.epsi}
\caption{
QMC result for
$P(r)$ with $U=-t$. $n=1$ ($\bigcirc$), $n=1/3$ ($\Box$).
}
  \end{center}
\label{fig2}
\end{figure}

In conclusion, we have shown that the effect of the spin gap
can be clearly seen in the correlation function of finite size systems 
only when the size of the gap is sufficiently large compared 
to the discreteness of the 
one-body energy levels. Combining the results 
for the Hubbard ladders and for the $t-t^{\prime}-U$ model, 
we believe that this is the case in general: 
A criterion for reproducing the correct form of the correlation functions 
is to make $\Delta_S/\Delta\varepsilon$ large.  
This implies that to reproduce the correct asymptotic behavior 
of the correlation functions in models having small spin gaps,
very large systems are required.  
An alternative way is to tune, 
if possible, the values of the parameters so 
that $\Delta\varepsilon$ becomes small
around the Fermi level, as we have done for ladders.

One final point is, 
since $\Delta\varepsilon$ is defined as 
the level offset at $U=0$, the quantity can become ill-defined 
for large interactions, where we have to consider 
that the quantity should be renormalized.

\section{Summary}

In the present paper, we have studied the pairing 
correlation in the Hubbard ladder with 
two(even) or  three(odd) number of legs. 
This has been motivated from a conjecture due to Rice {\it et al.} 
that an even-numbered ladder should exhibit 
dominance of the interchain singlet pairing correlation as expected 
from the persistent spin gap away from half-filling. 
Naively, one can then expect that an odd-numbered ladder 
should not exhibit dominance of the pairing correlation
reflecting the presence of gapless spin excitations.

We have first considered the two-leg Hubbard ladder model.
In the weak-coupling theory, in which the interactions 
are treated with the perturbative renormalization-group method, 
a d-wave like pairing correlation becomes dominant reflecting 
a spin gap in the two-leg Hubbard ladder. 
The relevant scattering processes are the pair-tunneling 
process across the bonding and the anti-bonding bands 
and the backward-scattering process within each band. 
The pair-tunneling process is reminiscent of the 
Suhl-Kondo mechanism for superconductivity in the transition
metals with a two-band structure. However, 
the weak-coupling theory is correct only for infinitesimally small 
interactions in principle. Thus the calculation for finite interaction $U$
is needed, but existing numerical calculations for finite $U$
have been controversial.

In Sec.II, we have applied the projector Monte Carlo method to investigate
the pairing correlation function in the ground state for finite $U$.
We conclude that the discreteness of energy levels in finite systems affects 
the pairing correlation enormously, where 
the enhanced pairing correlation is indeed detected for intermediate 
interaction strengths 
if we tune the parameters so as to align 
the discrete energy levels of
bonding and anti-bonding bands at the Fermi level 
in order to mimic the thermodynamic limit. 
The enhancement of the pairing correlation in the $U=2t$ case 
is smaller than that in the $U=t$ case. 
This is consistent with the weak-coupling 
theory in which the pairing correlation decays as $r^{-1/(2K_{\rho})}$
($K_\rho$ is the critical exponent of the gapless charge mode) at 
long distances and $K_\rho$ is a {\it decreasing} function of $U$.

In the cases where interband or intraband Umklapp process is possible, 
the pairing correlation is not enhanced. 
This result is also consistent with  
the weak-coupling theory. 

We then moved onto the correlation functions 
in the three-leg Hubbard ladder model in Sec.III. 
Whether the above `even-odd' conjecture 
holds for the simplest-odd ladder (i.e. the three-leg ladder) 
with a plural number of charge and spin modes
is an important problem. This has remained an open question,  
since there had been no results for the
pairing correlation function in the three-leg $t-J$ 
or Hubbard ladder models. 

A key is the coexistence of gapless and gapful spin excitations 
in the doped three-leg Hubbard ladder. 
This has been analytically shown from the correlation functions 
starting from the phase diagram obtained by Arrigoni\cite{Arrigoni}, 
who enumerated  the numbers of the gapless charge and spin modes 
with the perturbative renormalization-group technique 
in the weak-coupling limit. 
If we turn to the correlation functions, 
we have found that the coexisting gapful and gapless modes give rise to 
a peculiar situation where a specific pairing {\it across} the central and 
edge chains (roughly a d-wave pairing) is dominant, 
while the $2k_F$ SDW on the edge chains 
simultaneously shows a subdominant but still 
long-tailed (power-law) decay  
associated with the gapless spin mode.
The relevant scattering processes are the 
pair-tunneling process between the top and the bottom bands 
and the backward-scattering process within the top and the bottom bands.
The situation is rather similar to the two leg case where
the pair-tunneling processes play an important role for the 
enhanced pairing correlation. 
Schulz\cite{Schulz3} has independently obtained results 
for both the SDW and the pairing correlations which are 
similar to those given in Sec.III.

However, as discussed in the two-leg case, 
it is not clear whether the weak-coupling 
results might be applicable only to 
infinitesimally small interaction strengths. 
In Sec.IV, we have thus checked the pairing 
correlation in the three-leg 
Hubbard ladder with the QMC method tuning 
the parameters so as to align the discrete energy levels of 
the first and third bands at the Fermi level as in the two-leg case. 
The enhanced pairing correlation is indeed detected even 
for intermediate interaction strengths in the three-leg ladder. 
The enhancement of the pairing correlation in the $U=2t$ case is 
smaller than that in the $U=t$ case as in the two-leg case in Sec.II.
This result is consistent with the weak-coupling theory in Sec.III 
in a similar reason with that in the two-leg case. Namely, the 
exponent of the pairing correlation is a decreasing function of 
$K_{\rho}$'s which should decrease with $U$.

Various effects of the Umklapp processes have also been discussed.  
Especially, it is found that the enhancement of the pairing correlation
is not affected by the intraband Umklapp process within the second band 
which does not involve the pairing order parameter. 
The key message obtained 
in the present paper, is that the 
dominance of the pairing correlation only requires the existence 
of gap(s) in not all but {\it some} of the spin modes. 
This is independent to 
whether the number of legs is even(two) or  odd(three). 

There are still important open questions. 
One is how the Hubbard model can possibly be related to real systems such 
as the cuprates. Specifically, 
One should study the two-leg and/or three-leg 
Hubbard ladder for larger $U$, 
where the dominance of the pairing correlation 
might be lost. 
Furthermore, the three-leg $t-J$ model should be studied to be compared 
with the Hubbard ladder. It would also be interesting to further investigate
ladder systems with larger numbers of legs. 
The Hubbard ladders with more large number of legs is also of interest, 
since it may have a key point to understand the two-dimensional systems. 
More recently, 
Lin {\it et al.}\cite{Lin} have examined the weak-coupling phase diagram 
for the four-leg Hubbard ladder where the pairing correlation 
is the most dominant in wide parameter region.

Finally, in Sec.V, a QMC study for the pairing correlation in 
the 1D attractive Hubbard model is given. 
The model can be exactly solved by Bethe Ansatz, so that 
we know the exact form of the correlation functions at long distances 
and the values of the spin gaps. 
A QMC study has shown that the effect of the spin gap 
can be clearly seen in the pairing correlation function of 
finite size systems and the behavior at long distances is 
consistent with the exact result 
only when the size of the gap is sufficiently 
large compared to the discreteness of the one-body energy levels. 
The result is a circumstantial evidence 
to justify the enhancement of 
the pairing correlation in Sec.II and Sec.IV and 
combining this result, we believe that 
this is the case in general for finite size systems.

We are grateful to Professor H.J. Schulz for directing our attention 
to refs.\cite{SchulzAF} and \cite{Schulz3}. 
We wish to thank E. Arrigoni for sending us 
his work prior to publication. 
We also acknowledge helpful discussions with 
Professors M. Ogata, Y. Takada, H. Fukuyama, M. Imada, 
A. Fujimori, N. M{\^o}ri, M. Azuma, Z. Hiroi, and J. Akimitsu. 
Numerical calculations were performed on a 
FACOM VPP 500/40 at the Supercomputer Center, Institute for 
Solid State Physics, the University of Tokyo and on a HITAC S3800/280
at the Computer Center of the University of Tokyo. 
For the latter facility we thank Professor Y. Kanada for 
support in a Project for Vectorized Super Computing. 
This work was also supported in part by a Grant-in-Aid for 
Scientific Research from the Ministry of Education, Science, 
Sports, and Culture of Japan. One of the authors (T.K.) 
acknowledges the Japan Society for the Promotion of Science for 
a fellowship.
 
\appendix
\section{Calculation Method in Sec.III} 
\subsection{Derivation of eq.(3.4)}
Here we derive eq.(3.4) in the standard bosonization method.
The intra- and the inter-band forward-scattering terms, which 
can be diagonally treated in the phase Hamiltonian
as we will see in the following, are 
produced from the intrachain forward-scattering terms using eq.(3.3) 
as follows:
\begin{eqnarray}
\lefteqn{\frac{2\pi v_F g}{L}\sum_{k_1,k_2,q}\sum_{\mu,\sigma,\sigma'}
c_{\mu,+,k_1,\sigma}^{\dagger}
c_{\mu,-,k_2,\sigma'}^{\dagger}
c_{\mu,-,k_2+q,\sigma'}
c_{\mu,+,k_1-q,\sigma}}
\nonumber\\
&&\equiv H_{{\rm f}}+\mbox{pair-tunneling terms}, \nonumber\\
&&=\frac{\pi v_F g}{4}
\sum_{k_1,k_2,q}\sum_{\sigma,\sigma'}
[3
a_{1,+,k_1,\sigma}^{\dagger}
a_{1,-,k_2,\sigma'}^{\dagger}
a_{1,-,k_2+q,\sigma'}a_{1,+,k_1-q,\sigma}
\nonumber\\
&&+4
a_{2,+,k_1,\sigma}^{\dagger}a_{2,-,k_2,\sigma'}^{\dagger}
a_{2,-,k_2+q,\sigma'}a_{2,+,k_1-q,\sigma}\nonumber\\
&&+3
a_{3,+,k_1,\sigma}^{\dagger}a_{3,-,k_2,\sigma'}^{\dagger}
a_{3,-,k_2+q,\sigma'}a_{3,+,k_1-q,\sigma}\nonumber\\
&&+2(a_{1,+,k_1,\sigma}^{\dagger}a_{2,-,k_2,\sigma'}^{\dagger}
a_{2,-,k_2+q,\sigma'}a_{1,+,k_1-q,\sigma}\nonumber\\
&&+a_{3,+,k_1,\sigma}^{\dagger}a_{2,-,k_2,\sigma'}^{\dagger}
a_{2,-,k_2+q,\sigma'}a_{3,+,k_1-q,\sigma}\nonumber\\
&&+
a_{2,+,k_1,\sigma}^{\dagger}a_{1,-,k_2,\sigma'}^{\dagger}
a_{1,-,k_2+q,\sigma'}a_{2,+,k_1-q,\sigma}\nonumber\\
&&+a_{2,+,k_1,\sigma}^{\dagger}a_{3,-,k_2,\sigma'}^{\dagger}
a_{3,-,k_2+q,\sigma'}a_{2,+,k_1-q,\sigma})\nonumber\\
&&+
3(a_{1,+,k_1,\sigma}^{\dagger}a_{3,-,k_2,\sigma'}^{\dagger}
a_{3,-,k_2+q,\sigma'}a_{1,+,k_1-q,\sigma}\nonumber\\
&&+a_{3,+,k_1,\sigma}^{\dagger}a_{1,-,k_2,\sigma'}^{\dagger}
a_{1,-,k_2+q,\sigma'}a_{3,+,k_1-q,\sigma})
]
\nonumber\\
&&+\mbox{pair-tunneling terms}.
\end{eqnarray}
Here $H_{\rm f}$ consists of the forward-scattering processes in the
band description and the electron operator with index $+(-)$ 
belongs to the right-(left-) going branch. 
We prepare the following bosonic operators as in the usual 
single-chain case.
\begin{equation}\alpha_{i,k} = \left\{
\begin{array}{rl}
&\Big(\frac{\pi}{kL}\Big)^{1/2}\sum_{i,p,\sigma} 
a_{i,+,p-k,\sigma}^{\dagger}
a_{i,+,p,\sigma}
\hspace{3.3mm} \quad \mbox{for $k>0$},\\
&\Big(\frac{\pi}{|k|L}\Big)^{1/2}\sum_{i,p,\sigma} 
a_{i,-,p+|k|,\sigma}^{\dagger}
a_{i,-,p,\sigma}
\quad \mbox{for $k<0$},
\end{array}\right. 
\end{equation}
\begin{equation}\beta_{i,k} = \left\{
\begin{array}{rl}
&\Big(\frac{\pi}{kL}\Big)^{1/2}\sum_{i,p,\sigma} 
\sigma a_{i,+,p-k,\sigma}^{\dagger}
a_{i,+,p,\sigma}
\hspace{3.3mm} \quad \mbox{for $k>0$},\\
&\Big(\frac{\pi}{|k|L}\Big)^{1/2}\sum_{i,p,\sigma} 
\sigma a_{i,-,p+|k|,\sigma}^{\dagger}
a_{i,-,p,\sigma}
\quad \mbox{for $k<0$}. 
\end{array}\right. 
\end{equation}
$\alpha_{i,k}(\beta_{i,k})$ corresponds to the charge(spin)-density
excitation in band $i$. 
Note that, $\alpha_{i,k}$ and $\beta_{i,k}$ obey the boson commutation
relation:
\begin{eqnarray}
[\alpha_{i,k},\alpha_{i',k'}]=[\beta_{i,k},\beta_{i',k'}]
=[\alpha_{i,k},\beta_{i',k'}]=[\alpha_{i,k},\beta_{i',k'}^{\dagger}]=0,
\nonumber
\end{eqnarray}
\begin{eqnarray}
[\alpha_{i,k},\alpha_{i',k'}^{\dagger}]=[\beta_{i,k},\beta_{i',k'}^{\dagger}]
=\delta_{i,i'}\delta_{k,k'}.
\end{eqnarray}
$H_{{\rm f}}$ is expressed in terms of only $\alpha_{i,k}$ as
\begin{eqnarray}
H_{{\rm f}}&=&\frac{v_F g}{4}
\sum_{k>0}k
[3(
\alpha_{1,k}^{\dagger}\alpha_{1,-k}^{\dagger}+
\alpha_{3,k}^{\dagger}\alpha_{3,-k}^{\dagger}\nonumber\\
&+&
\alpha_{1,k}\alpha_{1,-k}+
\alpha_{3,k}\alpha_{3,-k})
+4(
\alpha_{2,k}^{\dagger}\alpha_{2,-k}^{\dagger}+
\alpha_{2,k}\alpha_{2,-k})\nonumber\\
&+&2(
\alpha_{1,k}^{\dagger}\alpha_{2,-k}^{\dagger}+
\alpha_{1,k}\alpha_{2,-k}+
\alpha_{3,k}^{\dagger}\alpha_{2,-k}^{\dagger}+
\alpha_{3,k}\alpha_{2,-k}\nonumber\\
&+&
\alpha_{1,-k}^{\dagger}\alpha_{2,k}^{\dagger}+
\alpha_{1,-k}\alpha_{2,k}
+
\alpha_{3,-k}^{\dagger}\alpha_{2,k}^{\dagger}
+\alpha_{3,-k}\alpha_{2,k})\nonumber\\
&+&3(
\alpha_{1,k}^{\dagger}\alpha_{3,-k}^{\dagger}+
\alpha_{1,k}\alpha_{3,-k}+
\alpha_{3,k}^{\dagger}\alpha_{1,-k}^{\dagger}+
\alpha_{3,k}\alpha_{1,-k})
].\nonumber\\
\
\end{eqnarray}
Furthermore we can rewrite the non-interacting part of the
Hamiltonian $H_0$ as
\begin{eqnarray}
H_0 =
v_F\sum_{i,p}|p|
\alpha_{i,p}^{\dagger}\alpha_{i,p}+
v_F\sum_{i,p}|p|
\beta_{i,p}^{\dagger}\beta_{i,p}.
\end{eqnarray}
Now we introduce the phase variables as in the single chain case
by the following equations:
\begin{eqnarray}
\theta_{i\pm}&=&i\sum_{k>0}\sqrt{\frac{\pi}{Lk}}
e^{-\Lambda k/2}\nonumber\\
&&\times
[e^{-ikx}(\alpha_{i,k}^{\dagger}\pm\alpha_{i,-k})
-e^{ikx}(\alpha_{i,k}\pm\alpha_{i,-k}^{\dagger})],\\
\phi_{i\pm}&=&i\sum_{k>0}\sqrt{\frac{\pi}{Lk}}
e^{-\Lambda k/2}\nonumber\\
&&\times[e^{-ikx}(\beta_{i,k}^{\dagger}\pm\beta_{i,-k})
-e^{ikx}(\beta_{i,k}\pm\beta_{i,-k}^{\dagger})].
\end{eqnarray}
Here the phase variable $\phi_{i+}(\theta_{i+})$
can be regarded as the phase of the spin(charge)-density wave,
while $\phi_{i-}(\theta_{i-})$ is the field dual to 
$\phi_{i+}(\theta_{i+})$. 

From (B.7) and (B.8), we can write the diagonal part of the 
Hamiltonian, $H_{\rm d}=H_0+H_{\rm f}$, in
terms of the phase variables as
\begin{eqnarray}
\lefteqn
H_{\rm d}=&&H_{\rm spin}+H_{\rm charge}, \nonumber\\
H_{\rm spin}=&&\sum_i\frac{v_F}{4\pi}\int\ dx
[(\partial_x\phi_{i+})^2+(\partial_x\phi_{i-})^2], \\
H_{\rm charge}&&\nonumber\\
=\frac{v_F}{4\pi}\int\ dx&&\Big[
\Big(1+\frac{3}{4}g\Big) (\partial_x\theta_{1+})^2+
\Big(1-\frac{3}{4}g\Big)(\partial_x\theta_{1-})^2 
\Big]\nonumber\\
+\frac{v_F}{4\pi}\int\ dx&&\Big[
\Big(1+g\Big) (\partial_x\theta_{2+})^2+
\Big(1-g\Big)(\partial_x\theta_{2-})^2 
\Big]\nonumber\\
+\frac{v_F}{4\pi}\int\ dx&&\Big[
\Big(1+\frac{3}{4}g\Big) (\partial_x\theta_{3+})^2\nonumber\\
&&+\Big(1-\frac{3}{4}g\Big)(\partial_x\theta_{3-})^2 
\Big]\\
+\frac{v_F g}{4\pi}\int\ dx&&\Big[
(\partial_x\theta_{1+})(\partial_x\theta_{2+})
-(\partial_x\theta_{1-})(\partial_x\theta_{2-})\nonumber\\
&&+(\partial_x\theta_{3+})(\partial_x\theta_{2+})
-(\partial_x\theta_{3-})(\partial_x\theta_{2-})
\Big]\nonumber\\
+\frac{3 v_F g}{8\pi}\int\ dx&&\Big[
(\partial_x\theta_{1+})(\partial_x\theta_{3+})
-(\partial_x\theta_{1-})(\partial_x\theta_{3-})
\Big].\nonumber
\end{eqnarray}
Thus $H_{\rm d}$ is separated to both the spin-part $H_{\rm spin}$
and the charge-part $H_{\rm charge}$.
$H_{\rm charge}$ is also diagonalized 
by using eq.(3.5), while $H_{\rm spin}$ is already diagonalized. 
As a result, eq.(3.4) is easily obtained.

\subsection{Calculation of Correlation Functions}
Here we explain the method to derive the correlation functions.
As examples, we here calculate the correlation functions
of the intrachain singlet pairing in the edge chains and of the
singlet pairing across the central and the edge chains.
As stated in the text, the relevant scattering processes,
are the pair-tunneling process 
between the first and the third bands and 
the backward-scattering process within the 
first or the third band. 

The pair-tunneling process is expressed in terms of the 
phase variables as follows:
\begin{eqnarray}
&&\psi_{3+\uparrow}^{\dagger}\psi_{3-\downarrow}^{\dagger}
\psi_{1-\downarrow}\psi_{1+\uparrow}+
\psi_{3+\downarrow}^{\dagger}\psi_{3-\uparrow}^{\dagger}
\psi_{1-\uparrow}\psi_{1+\downarrow}\nonumber\\
&+&\psi_{3+\downarrow}^{\dagger}\psi_{3-\uparrow}^{\dagger}
\psi_{1+\uparrow}\psi_{1-\downarrow}+
\psi_{3+\uparrow}^{\dagger}\psi_{3-\downarrow}^{\dagger}
\psi_{1+\downarrow}\psi_{1-\uparrow}+{\rm h.c.}\nonumber\\
&\propto&
\eta_{3+\uparrow}\eta_{3-\downarrow}\eta_{1-\downarrow}\eta_{1+\uparrow}
{\rm exp}[i(\theta_{1-}-\theta_{3-}+\phi_{1+}-\phi_{3+})]\nonumber\\
&+&\eta_{3+\downarrow}\eta_{3-\uparrow}\eta_{1-\uparrow}\eta_{1+\downarrow}
{\rm exp}[i(\theta_{1-}-\theta_{3-}-\phi_{1+}+\phi_{3+})]\label{siki}\\
&+&\eta_{3+\downarrow}\eta_{3-\uparrow}\eta_{1+\uparrow}\eta_{1-\downarrow}
{\rm exp}[i(\theta_{1-}-\theta_{3-}+\phi_{1+}+\phi_{3+})]\nonumber\\
&+&\eta_{3+\uparrow}\eta_{3-\downarrow}\eta_{1+\downarrow}\eta_{1-\uparrow}
{\rm exp}[i(\theta_{1-}-\theta_{3-}-\phi_{1+}-\phi_{3+})]
+{\rm h.c.},\nonumber\\
&=&2{\rm exp}[i(\theta_{1-}-\theta_{3-})]
[{\rm cos}(\phi_{1+}-\phi_{3+})-{\rm cos}(\phi_{1+}+\phi_{3+})
]+{\rm h.c.},
\nonumber\\
&=&8{\rm cos}(\sqrt{2}\chi_{1-})
{\rm sin}(\phi_{1+}){\rm sin}(\phi_{3+}).\nonumber
\end{eqnarray}
Here we have defined the product of the $U$ operators\cite{Hal1D1} as
$\eta_{i+\uparrow}\eta_{i-\downarrow}=\eta_{i+\downarrow}\eta_{i-\uparrow}$,
but this convention does not affect the correlation functions
as in the two-leg case\cite{Fabrizio,Schulz2}.

The backward-scattering process within band $i$ 
is also expressed in terms of the phase variables
 through eq.(3.6) as
\begin{eqnarray}
&&\psi_{i+\uparrow}^{\dagger}\psi_{i-\downarrow}^{\dagger}
\psi_{i+\downarrow}\psi_{i-\uparrow}+{\rm h.c.}\nonumber\\
&=&\eta_{i+\uparrow}\eta_{i-\downarrow}
\eta_{i+\downarrow}\eta_{i-\uparrow}
{\rm exp}[-2i\phi_{i+}]
+{\rm h.c.},
\\
&=&-2{\rm cos}(2\phi_{i+}).\nonumber
\end{eqnarray}
This and eq.(\ref{siki}) give the eq.(3.7) in the text. 
In the beginning we calculate the correlation function of the intrachain
pairing in an edge chain ($\alpha$ chain).
The order parameter is expressed in the band description as
\begin{eqnarray}
O_{\rm intraSS}
&\equiv&
\psi_{\alpha+\uparrow}\psi_{\alpha-\downarrow}\nonumber\\
&\sim&\frac{1}{4}[\psi_{1+\uparrow}\psi_{1-\downarrow}
+\psi_{3+\uparrow}\psi_{3-\downarrow}
+2\psi_{2+\uparrow}\psi_{2-\downarrow}],
\end{eqnarray}
where we have picked up only the 
two-particle operators 
whose correlations show power-law decay at long distances. 
The correlations of the other two-particle operators 
correlation decay exponentially due to the gapful field(s).
We can rewrite the above equation in the phase variables as
\begin{eqnarray}
&&\psi_{1+\uparrow}\psi_{1-\downarrow}
+2\psi_{2+\uparrow}\psi_{2-\downarrow}
+\psi_{3+\uparrow}\psi_{3-\downarrow}\nonumber\\
&\propto&\eta_{1+\uparrow}\eta_{1-\downarrow}
{\rm exp}[i(\theta_{1-}+\phi_{1+})]
+\eta_{3+\uparrow}\eta_{3-\downarrow}
{\rm exp}[i(\theta_{3-}+\phi_{3+})]\nonumber\\
&+&2\eta_{2+\uparrow}\eta_{2-\downarrow}
{\rm exp}[i(\theta_{2-}+\phi_{2+})],\nonumber\\
&=&
i{\rm exp}
\Big\{i[(\frac{1}{\sqrt{2}}\chi_{1-}+\frac{1}{\sqrt{3}}\chi_{2-}+\frac{1}{\sqrt{6}}\chi_{3-})
+\phi_{1+}]\Big\}\nonumber\\
&&+i{\rm exp}
\Big\{i[(-\frac{1}{\sqrt{2}}\chi_{1-}+\frac{1}{\sqrt{3}}\chi_{2-}+\frac{1}{\sqrt{6}}\chi_{3-})
+\phi_{3+}]\Big\}\nonumber\\
&&+2i{\rm exp}[i(\theta_{2-}+\phi_{2+})],\nonumber\\
&=&i{\rm exp}
\Big\{i[(\frac{\pi}{2}+\frac{1}{\sqrt{3}}\chi_{2-}+\frac{1}{\sqrt{6}}\chi_{3-})+\frac{\pi}{2}]\Big\}\nonumber\\
&&+i{\rm exp}
\Big\{i[(-\frac{\pi}{2}+\frac{1}{\sqrt{3}}\chi_{2-}+\frac{1}{\sqrt{6}}\chi_{3-})+\frac{\pi}{2}]\Big\}\nonumber\\
&&+2i{\rm exp}[i(\theta_{2-}-\phi_{2+})],\\
&=&2i{\rm exp}[i(\theta_{2-}-\phi_{2+})].\nonumber
\end{eqnarray}
Here we have fixed $\phi_{1+}=\phi_{3+}=\pi/2$ and $\chi_{1-}=\pi/\sqrt{2}$ as discussed in the text and the terms containing t
he gapful fields are 
canceled out.
Now we can calculate the correlation function:
\begin{eqnarray}
&&\langle O_{{\rm intraSS}}(x)O_{{\rm intraSS}}(0)\rangle
\nonumber\\
&\propto&
\langle 
{\rm exp}[i(\theta_{2-}(x)-\phi_{2+}(x))]
{\rm exp}[i(\theta_{2-}(0)-\phi_{2+}(0))] 
\rangle,
\nonumber\\
&=&{\rm exp}
[-\frac{1}{2} \{ \langle(\theta_{2-}(x)-\theta_{2-}(0))^2 \rangle
+\langle (\phi_{2+}(x)-\phi_{2+}(0))^2 \rangle \} ],\nonumber\\
&=&{\rm exp}
[-\frac{2\pi}{3L} (\frac{1}{K_{\rho 2}^*}+\frac{2}{K_{\rho 3}^*}+3)
\sum_{k>0} \frac{e^{-\Lambda k}}{k}(1-{\rm cos}kx) ],
\\
&=&{\rm exp}
[-\frac{1}{6} (\frac{1}{K_{\rho 2}^*}+\frac{2}{K_{\rho 3}^*}+3) 
{\rm log} (1+ \frac{x^2}{\Lambda^2}) ],\nonumber\\
&\sim& x^{-\frac{1}{3}(\frac{1}{K_{\rho 2}^*}+\frac{2}{K_{\rho 3}^*})-1}.\nonumber
\end{eqnarray}
Now we calculate the correlation function of the singlet pairing across the 
central and the edge chains. The order parameter is expressed as
\begin{eqnarray}
O_{\rm CESS}&=&
(\psi_{\alpha+\uparrow}+\psi_{\gamma+\uparrow})\psi_{\beta-\downarrow}
-(\psi_{\alpha+\downarrow}+\psi_{\gamma+\downarrow})
\psi_{\beta-\uparrow},\nonumber\\
&\sim& 
\psi_{1+\uparrow}\psi_{1-\downarrow}-\psi_{3+\uparrow}\psi_{3-\downarrow}
\nonumber\\
&&-(\psi_{1+\downarrow}\psi_{1-\uparrow}-\psi_{3+\downarrow}\psi_{3-\uparrow}).
\end{eqnarray}
Here again we pick up only the two-particle operators 
whose correlations show power-law decay. In terms of the phase variables, 
the order parameter can be rewritten as
\begin{eqnarray}
&&\psi_{1+\uparrow}\psi_{1-\downarrow}-\psi_{3+\uparrow}\psi_{3-\downarrow}
-\psi_{1+\downarrow}\psi_{1-\uparrow}+\psi_{3+\downarrow}\psi_{3-\uparrow}
\nonumber\\
&\propto&
\eta_{1+\uparrow}\eta_{1-\downarrow}
{\rm exp}[i(\theta_{1-}+\phi_{1+})]
-\eta_{3+\uparrow}\eta_{3-\downarrow}
{\rm exp}[i(\theta_{3-}+\phi_{3+})]\nonumber\\
&-&\eta_{1+\downarrow}\eta_{1-\uparrow}
{\rm exp}[i(\theta_{1-}-\phi_{1+})] 
+\eta_{3+\downarrow}\eta_{3-\uparrow}
{\rm exp}[i(\theta_{3-}-\phi_{3+})],\nonumber\\
&=&i[
{\rm exp}[i\{(\frac{1}{\sqrt{2}}\frac{\pi}{\sqrt{2}}+\frac{1}{\sqrt{3}}\chi_{2-}+
\frac{1}{\sqrt{6}}\chi_{3-})+\frac{\pi}{2}\}]\nonumber\\
&&-{\rm exp}[i\{(-\frac{1}{\sqrt{2}}\frac{\pi}{\sqrt{2}}+\frac{1}{\sqrt{3}}\chi_{2-}+\frac{1}{\sqrt{6}}\chi_{3-})+\frac{\pi}{2}
\}]\\
&&-{\rm exp}[i\{(\frac{1}{\sqrt{2}}\frac{\pi}{\sqrt{2}}+\frac{1}{\sqrt{3}}\chi_{2-}
\frac{1}{\sqrt{6}}\chi_{3-})-\frac{\pi}{2}\}]\nonumber\\
&&+{\rm exp}
[i\{(-\frac{1}{\sqrt{2}}\frac{\pi}{\sqrt{2}}+\frac{1}{\sqrt{3}}\chi_{2-}+
\frac{1}{\sqrt{6}}\chi_{3-})-\frac{\pi}{2}\}],\nonumber\\
&=&-4i
{\rm exp}[i(\frac{1}{\sqrt{3}}\chi_{2-}+\frac{1}{\sqrt{6}}\chi_{3-})]\nonumber.
\end{eqnarray}
Calculation of the interchain pairing correlation function is 
quite similar to that of the intrachain pairing correlation.
\begin{eqnarray}
&&\langle O_{\rm CESS}(x)O_{\rm CESS}(0)\rangle\nonumber\\
&\propto&\langle {\rm exp}[i(\frac{1}{\sqrt{3}}\chi_{2-}(x)+\frac{1}{\sqrt{6}}
\chi_{3-}(x))]\nonumber\\
&&\times
{\rm exp}[i(\frac{1}{\sqrt{3}}\chi_{2-}(0)+\frac{1}{\sqrt{6}}\chi_{3-}(0))]
\rangle\nonumber,\\
&=&{\rm exp}
[-\frac{1}{2}\{\frac{1}{3}\langle(\chi_{2-}(x)-\chi_{2-}(0))^2\rangle
\nonumber\\
&&+\frac{1}{6}\langle(\chi_{3-}(x)-\chi_{3-}(0))^2\rangle\}],\\
&\sim&x^{-\frac{1}{3}(\frac{1}{K_{\rho2}^*}+\frac{1}{2K_{\rho3}^*})}.
\nonumber
\end{eqnarray}
From above calculations, we can see that the interchain pairing exploits 
the charge gap and the spin gaps to reduce the exponent of the correlation
function, in contrast to the intrachain pairing.


\begin{references}
\bibitem{*}*Present address: Department of Physical Electrics, 
Hiroshima University, Higashi-hiroshima 739, Japan. 
\bibitem{DagRice} See for a review, E. Dagotto and T.M. Rice, 
Science {\bf 271} 618 
(1996) and references therein. 
\bibitem{SchulzAF} H.J. Schulz, Phys. Rev. B {\bf 34}, 6372 (1986).
\bibitem{Haldane} F.D.M. Haldane, Phys. Lett. A {\bf 93}, 464 (1993).
\bibitem{Nishiyama} Y. Nishiyama, N. Hatano, and M. Suzuki,
J. Phys. Soc. Jpn {\bf 64}, 1967 (1994).
\bibitem{Rice} T.M. Rice, S. Gopalan, and M. Sigrist, 
Europhys. Lett. {\bf 23} 445 (1993); Physica B {\bf 199 \& 200}, 378 (1992).
\bibitem{Dagotto} E. Dagotto, J. Riera, and D.J. Scalapino, 
Phys. Rev. B {\bf 45}, 5744 (1992).
\bibitem{White} S.R. White, R.M. Noack, and D.J. Scalapino,
Phys. Rev. Lett {\bf 73}, 886 (1994).
\bibitem{Greven} M. Greven, R.J. Birgeneau, and U.-J. Wiese, 
Phys. Rev. Lett. {\bf 77}, 1865 (1996).
\bibitem{Poilblanc3} D. Poilblanc, H. Tsunetsugu, and T.M. Rice,
Phys. Rev. B {\bf 50}, 6511 (1994).
\bibitem{Hatano} N. Hatano and Y. Nishiyama, 
J. Phys. A {\bf 28}, 3911 (1995). 
\bibitem{Azuma} M. Azuma {\it et al.}, 
Phys. Rev. Lett. {\bf 73}, 3463 (1994). 
\bibitem{Ishida1} K. Ishida {\it et al.},
Phys. Rev. B {\bf 53}, 
2827 (1996). 
\bibitem{Kojima} K. Kojima {\it et al.}, 
Phys. Rev. Lett. {\bf 74}, 2812 (1995). 
\bibitem{Hiroi2} Z. Hiroi and M. Takano,
Nature {\bf 377}, 41 (1995).
\bibitem{Sigrist} M. Sigrist, T.M. Rice, and F.C. Zhang, 
Phys. Rev. B {\bf 49}, 12058 (1994).
\bibitem{Tsunetsugu} H. Tsunetsugu, M. Troyer, 
and T.M. Rice, Phys. Rev. B {\bf 49}, 16078 (1994);
{\it ibid} {\bf 51}, 16456 (1995).
\bibitem{Hayward1}
C.A. Hayward {\it et al.}, 
Phys. Rev. Lett. {\bf 75}, 926 (1995).
\bibitem{Hayward2}
C.A. Hayward and D. Poilblanc,
Phys. Rev. B {\bf 53}, 11721 (1996).
\bibitem{Hayward3} C.A. Hayward, D. Poilblanc, and D.J. Scalapino,
Phys. Rev. B {\bf 53}, R8863 (1996).
\bibitem{Sano}
K. Sano, J. Phys. Soc. Jpn. {\bf 65}, 1146 (1996).
\bibitem{Uehara}M. Uehara {\it et al.}, 
J. Phys. Soc. Jpn. {\bf 65}, 2764 (1996).
\bibitem{Review} 
See for reviews, J. S\'{o}lyom, Adv. Phys. {\bf 28}, 201 (1979); 
V.J. Emery, 
in {\it Highly Conducting One-Dimensional Solids}, 
ed. by J.T. Devreese {\it et al.} (Plenum, New York, 1979), p.247;
H. Fukuyama and H. Takayama, 
in {\it Electronic Properties of Inorganic 
Quasi-One Dimensional Compounds}, 
ed. by P. Mon\c{c}eau (D. Reidel, 1985), p.41.
\bibitem{weak} A. Luther and V.J. Emery,
Phys. Rev. Lett. {\bf 33}, 589 (1974);
P.A. Lee, Phys. Rev. Lett. {\bf 34}, 1247 (1975);
C.M. Varma and A. Zawadowski, 
Phys. Rev. B {\bf 32}, 7399 (1985);
K. Penc and J. S{\"o}lyom, Phys. Rev. B {\bf 44}, 12690 (1993); 
A.M. Finkel'stein and A.I. Larkin, 
Phys. Rev. B {\bf 47}, 10461 (1993).
\bibitem{Hirsch} J.E. Hirsch, Phys. Rev. B {\bf 31}, 4403 (1985).
\bibitem{MonteCarlo} M. Imada and Y. Hatsugai, 
J. Phys. Soc. Jpn. {\bf 58}, 3572 (1989);
N. Furukawa and M. Imada, J. Phys. Soc. Jpn. {\bf 61}, 3331 (1992);
S. Sorella {\it et al.}, 
Int. J. Mod. Phys. B {\bf 1}, 993 (1988);
S.R. White {\it et al.}, 
Phys. Rev. B {\bf 40}, 506 (1991);
W. von der Linden, I. Morenstern, and H. de Raedt, 
Phys. Rev. B {\bf 41}, 4669 (1990). 
\bibitem{Balents}L. Balents and M.P.A. Fisher, Phys. Rev. B {\bf 53}, 
12133 (1996).
\bibitem{Fabrizio}M. Fabrizio, Phys. Rev. B {\bf 48}, 15838 (1993);
M. Fabrizio, A. Parola, and E. Tosatti,
Phys. Rev. B {46}, 3159 (1992). 
\bibitem{Uchida} M. Knupfer {\it et al.}, Phys. Rev. B {\bf 55}, 7491 (1997).
\bibitem{Nagaosa}N. Nagaosa and M. Oshikawa, 
J. Phys. Soc. Jpn. {\bf 65}, 2241 (1996).
\bibitem{Schulz2}H.J. Schulz, Phys. Rev. B {\bf 53}, R2959 (1996).
\bibitem{Noack1}R.M. Noack, S.R. White, and D.J. Scalapino, 
Phys. Rev. Lett. {\bf 73}, 882 (1994). 
\bibitem{Noack3}R.M. Noack, S.R. White, and D.J. Scalapino,
unpublished (cond-mat/9601047).
\bibitem{Suhl} H. Suhl, B.T. Mattis, and L.R. Walker,
Phys. Rev. Lett. {\bf 3}, 552 (1959).
\bibitem{Kondo} J. Kondo,
Prog. Theor. Phys. {\bf 29}, 1 (1963).
\bibitem{Muttalib} K.A. Muttalib and V.J. Emery,
Phys. Rev. Lett. {\bf 57}, 1370 (1986).
\bibitem{Fabrizio2} M. Fabrizio, Phys. Rev. B {\bf 54}, 10054 (1996).
\bibitem{Arita} K. Kuroki, R. Arita, and H. Aoki, unpublished
(cond-mat/9702214)
\bibitem{Yamaji}
K. Yamaji and Y. Shimoi, Physica C {\bf 222}, 349 (1994); 
K. Yamaji, Y. Shimoi, and T. Yanagisawa, Physica C {\bf 235-240}, 
2221 (1994).
\bibitem{Asai}Y. Asai, Phys. Rev. B {\bf 52}, 10390 (1995).
\bibitem{WS} S.R. White and D.J. Scalapino, 
Phys. Rev. B {\bf 55}, 6504 (1997).
\bibitem{Kuroki}K. Kuroki, T. Kimura, and H. Aoki, 
Phys. Rev. B {\bf 54}, R15641 (1996).
\bibitem{Takashi1}T. Kimura, K. Kuroki, and H. Aoki, 
Phys. Rev. B {\bf 54}, R9608 (1996).
\bibitem{Arrigoni}E. Arrigoni, Phys. Lett. A {\bf 215}, 91 (1996);
Physica Status Solidi B {\bf 195}, 425 (1996). 
\bibitem{Schulz3} H.J. Schulz, unpublished (cond-mat/965075).
\bibitem{Takashi2} T. Kimura, K. Kuroki, and H. Aoki, 
J. Phys. Soc. Jpn. {\bf 66}, 1599 (1997).
\bibitem{Noack2}R.M. Noack, S.R. White, and D.J. Scalapino,
Europhys. Lett. {\bf 30}, 163 (1995); J. Low Temp. Phys. {\bf 99}, 593 (1995).
\bibitem{Kato} M. Kato, K. Shiota, and Y. Koike, 
Physica C {\bf 258} 284 (1996).
\bibitem{Bethe} H. Frahm and V.E. Korepin, 
Phys. Rev. B {\bf 42}, 10553 (1990);
N. Kawakami and S.K. Yang,
Phys. Lett. A {\bf 148}, 359 (1990); 
E.H. Lieb and F.Y. Wu,
Phys. Rev. Lett. {\bf 20}, 1445 (1990).
\bibitem{Hal1D1} F.D.M. Haldane, 
J. Phys. C {\bf 14}, 2585 (1981); 
Phys. Rev. Lett. {\bf 47}, 1840 (1981).
\bibitem{Lin} H. Lin, L. Balents, and M.P.A. Fisher, 
unpublished (cond-mat/9703055). 
\end{references}
\end{document}